\documentclass[aps,prd,preprint,floatfix,eqsecnum,noshowpacs,nofootinbib]{revtex4}
\usepackage{epsfig,amsmath,amssymb,amsfonts,bm}

\newcommand{\be}{\begin{equation}}
\newcommand{\ee}{\end{equation}}

\newcommand{\fet}[1]{\mbox{\boldmath $#1$}}
\newcommand{\beq}{\begin{equation}}
\newcommand{\eeq}{\end{equation}}
\newcommand{\beqa}{\begin{eqnarray}}
\newcommand{\eeqa}{\end{eqnarray}}
\newcommand{\nn}{\nonumber \\ }

\newcommand{\E}[1]{\ensuremath{\times10^{#1}}}
\newcommand{\Tref}[1]{Table~\ref{#1}}
\newcommand{\eref}[1]{(\ref{#1})}

\newcommand{\Cite}[1]{Ref.~\cite{#1}}
\newcommand{\Sec}[1]{Section~\ref{#1}}
\newcommand{\etal}{\textit{et~al.}}

\newcommand{\HeT}{\ensuremath{^3{\rm He}}}
\newcommand{\HeF}{\ensuremath{^4{\rm He}}}

\newcommand{\LiSix}{\ensuremath{^6{\rm Li}}}
\newcommand{\LiS}{\ensuremath{^7{\rm Li}}}
\newcommand{\BeS}{\ensuremath{^7{\rm Be}}}

\DeclareMathSymbol{\varGamma}{\mathord}{letters}{"00}

\begin{document}

\title{Varying the light quark mass: impact on the nuclear force and \\
Big Bang nucleosynthesis
}

\author{J. C. Berengut$^{(a)}$, E.~Epelbaum$^{(b)}$, V.~V.~Flambaum$^{(a)}$,
C.~Hanhart$^{(c),(d),(e)}$, 
U.-G.~Mei{\ss}ner$^{(c),(d),(e),(f),(g),(h)}$, J.~Nebreda$^{(i)}$,
and  J. R. Pel\'aez$^{(j)}$ }
\affiliation{$^{(a)}$School of Physics, University of New South Wales, Sydney, NSW 2052, Australia}
\affiliation{$^{(b)}$Institut f\"ur Theoretische Physik II, Ruhr-Universit\"at Bochum, D-44780 Bochum, Germany}
\affiliation{$^{(c)}$Institut f\"{u}r Kernphysik and J\"ulich Center for Hadron Physics, Forschungszentrum J\"{u}lich, D--52425 J\"{u}lich, Germany}
\affiliation{$^{(d)}$Institute for Advanced Simulation, Forschungszentrum
  J\"{u}lich, 
D--52425 J\"{u}lich, Germany}
\affiliation{$^{(e)}$ JARA --  Forces And Matter Experiments, Forschungszentrum J\"{u}lich, D--52425 J\"{u}lich, Germany}
\affiliation{$^{(f)}$  Helmholtz-Institut f\"ur Strahlen- und Kernphysik, Universit\"{a}t Bonn, D-53115 Bonn, Germany}
\affiliation{$^{(g)}$  Bethe Center for Theoretical Physics,  Universit\"{a}t Bonn, D-53115 Bonn, Germany}
\affiliation{$^{(h)}$ JARA --  High Performance Computing, Forschungszentrum J\"{u}lich, D--52425 J\"{u}lich, Germany}
\affiliation{$^{(i)}$Yukawa Institute for Theoretical Physics, Kyoto University, Kyoto 606-8502,
Japan}
\affiliation{$^{(j)}$ Dept. F\'{\i}sica Te\'orica II, Universidad Complutense, 28040 Madrid. Spain}


\begin{abstract}
The quark mass dependences of light element binding energies and nuclear scattering lengths
are derived using chiral perturbation theory
in combination with non--perturbative methods. In particular, we present new,
improved values for the quark mass dependence of meson resonances that enter
the nuclear force. A detailed analysis of the theoretical
uncertainties arising in this determination is presented. 
As an application we derive from a
comparison of observed and calculated primordial deuterium and helium
abundances a stringent limit on the variation of the light 
quark mass, $\delta m_q/m_q = 0.02\pm0.04$.
Inclusion of the neutron lifetime modification under the assumption of a
variation of the Higgs vacuum expectation value that translates into  changing quark, 
electron, and weak gauge boson masses, leads to a stronger limit,
$|\delta m_q/m_q| < 0.009$.
\end{abstract}

\maketitle

\section{Introduction}

The Standard Model is widely believed to be a low-energy manifestation of a 
more general theory that unifies the four fundamental forces of nature. 
Several candidate unified theories suggest that spatial and temporal variation 
of fundamental constants is a possibility, or even a necessity, in an
expanding Universe~(see, e.g.~the reviews~\cite{uzanrev,olive:2002tz}). Studies of 
Big Bang nucleosynthesis (BBN) provide a unique probe of the values of 
fundamental constants in the pre-recombination Universe.
A further motivation to consider the response of light
nuclei to changes in $m_q$, the light quark mass\footnote{Throughout most of
  this paper, we work in the
isospin limit $m_u = m_d$ and only consider the average light quark mass,
$m_q = (m_u+m_d)/2$; in addition the quark masses are studied in
units of a fixed $\Lambda_\textrm{QCD}$, for only the variation of dimensionless quantities
is meaningful. However, in Sec.~\ref{sec:tau} we address the constraints from
neutron $\beta$-decay which necessarily requires the inclusion of isospin violation
by strong and electromagnetic effects.}, is related to anthropic 
considerations~\cite{jaffe:2008gd,hogan:1999wh} that have
e.g. been used in the context of carbon production in hot stars
\cite{Livio,Oberhummer:2000zj,Epelbaum:2012iu} in order to understand how much 
fine-tuning is necessary amongst the fundamental parameters of the Standard
Model in order to allow life to emerge on earth. Only now -- based on methods as used here --
is one really able to study the explicit quark mass
dependence of the nuclear forces and nuclear properties and therefore their
impact on, e.g., nuclear abundances, because such issues can only be investigated
systematically and completely based on chiral effective field theories
or lattice simulations (or combinations thereof).

In addition, recent studies of quasar absorption spectra suggest a cosmological
gradient in the value of the fine-structure constant, $\alpha$, across the
universe~\cite{alphadipole}. The existence of this spatial variation could be
confirmed from complementary astrophysical studies such as Big Bang
nucleosynthesis~\cite{berengut11prd}. If the values of fundamental constants
were different in different regions of space at the time of nucleosynthesis,
this could be seen in the spatial distribution of primordial deuterium
abundances. Note that while BBN is relatively insensitive to
$\alpha$-variation~\cite{dent07prd}, the limits placed on quark mass variation
in this work can be related to the variation of $\alpha$ under a range of
unification models~\cite{GUT}. Indeed many of these grand unification theories
predict that relative variations in the strong force would be an order of
magnitude or two larger than those of the electroweak forces (for a simple
explanation of this see, e.g.,~\cite{victorandbob}).
This is also connected to anthropic questions, for if a spatial
variation of fundamental constants were to exist, we should not ask how
finely tuned the fundamental parameters are, but instead conclude that life emerged
in the region of the universe where the parameters allowed for it.

Relating the observed primordial abundances to the values of fundamental
constants at the time of Big Bang nucleosynthesis requires theoretical models 
for how the nuclear reaction rates depend on observable quantities such as
binding energies and scattering lengths, as well as a model for how those
quantities in turn depend on the fundamental constants. The former has been
dealt with previously in several works, see
e.g.~\cite{dent07prd,berengut10plb,civitarese10npa,coc12prd} 
and the references therein; in this paper we provide a response matrix based
on the method described in~\cite{berengut10plb} for some of these quantities 
which are of importance to the current work.
The second part of the problem, relating bulk nuclear quantities to values 
of fundamental constants, is the focus of this paper. 

Most of the previous studies in this context were performed on the basis of 
model-dependent estimates for
quark-mass dependences of nuclear 
properties~\cite{victoredward,craigs,craigs2,adelaide,flambaum03arxiv,flambaum04prd}. However, 
there are two theoretical tools available that allow, in principle, for 
a model-independent access to quark mass dependences. On the one hand
there is lattice QCD, on the other hand one has chiral perturbation theory (ChPT). The
former is a direct evaluation of QCD in Euclidean space-time and 
thus the quark mass dependence is explicit. In the latter case, the operator structure of 
quark mass terms is fixed by the QCD symmetries; in fact, ChPT is 
a faithful representation of the spontaneous and explicit chiral 
symmetry breaking of QCD~\cite{Leutwyler:1993iq}.
The strength parameters (usually called low-energy constants, LECs)
of those operators have to be fixed from other sources. 
Generally, this is done by comparison with experiment. However, for operators
that explicitly involve the quark mass, as is the case here, such a
determination is difficult since in nature the quark masses take definite
values. To determine the LECs of such operators, one can either
 fit to lattice data directly (see, e.g.,
\cite{brunsmeissner}, where the formalism is outlined for the $\rho$
meson) or from low-energy scattering data when using some unitarization
scheme in addition to ChPT (see, e.g., Refs.~\cite{su2iam,Pelaez:2010fj,su3iam}). 
It should be stressed, however, that in the latter case some model-dependence
is involved, since the quark mass terms of higher order than what was
put in from ChPT are not complete and depend on the scheme used~\cite{gassermeissner}.
In some cases, this model-dependence can be controlled, to some extent, by a comparison with lattice data.

In Ref.~\cite{platterbedaque}, an effective field theory
treatment of the impact of quark  mass variation on BBN was
presented  for the first time. In
this work the quark mass dependence of the $NN$ scattering lengths was used as
primary input. To constrain these, the results of
Ref.~\cite{chiralextrapol_bonn} were used, since at present the lattice is not
sufficiently accurate to provide precise values of these fundamental parameters. 
Central to the analysis of Ref.~\cite{chiralextrapol_bonn} was a naturalness 
assumption for the  quark
mass corrections to the leading quark mass independent contact interactions.
On the other hand, the same LEC was
allowed to vary in a different range in Ref.~\cite{chiralextrapol_seattle},
which led to quite different quark mass variations of the two-nucleon
properties. We remark, however, that the considerations in Ref.~\cite{chiralextrapol_bonn} 
were consistent with the earlier resonance saturation study of the leading
and next-to-leading order contact interactions performed in Ref.~\cite{ressat}.

 In this work we study systematically the impact of quark mass variations on
 two-nucleon observables based on a study of the quark mass dependences of
 mesons, since those are expected to give the most prominent
 contributions. In particular, if a strong quark mass dependence of, say, the
 potential part of the nucleon-nucleon ($NN$) interaction that comes from $\sigma$ exchange were
 present --- e.g. in Ref.~\cite{victoredward} a striking strange quark mass
 dependence of the $\sigma$ is conjectured --- it might, in the effective
 field theory where this field is integrated out, lead to an unnaturally
 enhanced LEC accompanying some contact interaction.  An example for such a
 pattern are some low-energy constants of dimension two that appear in $\pi N$
 scattering: when extracted from data in standard ChPT they appear to be
 unnaturally large, however, this can be understood phenomenologically by
 observing that they are mainly dominated by the exchange of $\Delta$
 isobars~\cite{ulfs}. Consequently, once the $\Delta$ contribution is subtracted
 the residual LECs get reduced significantly. Analogously one
 should expect that, once all meson exchange induced large quark mass effects
 are treated explicitly, the bulk of the quark mass dependence is
 included\footnote{In addition, we need to assume that there are no large quark masses
dependences coming from sources other than meson exchanges. In this sense 
the findings of Ref.~\cite{soto_new} are important for here it is demonstrated
that potentially large quark mass dependences induced by $\pi NN$ cuts~\cite{Mondejar:2006yu}
cancel, once final state interaction effects are considered explicitly.}.
 Still, such a procedure involves some modeling that induces some
 systematic uncertainty which is very difficult to specify.

Our main focus here are the sigma and the rho meson. Both appear as resonances
in the two--pion system. The cleanest way to connect their properties to the
$NN$ sector is via a dispersion integral of the Omn\`es type as used, e.g., in
Refs.~\cite{john1,john2}. Here, however, we use a method which is
technically easier to handle
and more transparent, although admittedly of lower theoretical rigour:
in Ref.~\cite{ressat} it was shown that the four nucleon operators of the $NN$
potential can be understood quantitatively in terms of the exchange of heavy
mesons in the sense of a resonance saturation. In that paper explicit
expressions are presented for this kind of matching. Thus we here use the
following
strategy: we determine the quark mass dependence of the light resonances
using the methods of Refs.~\cite{su2iam,Pelaez:2010fj,su3iam} which allows us to
predict the quark
mass  dependencies of the four--nucleon contact terms using the
expressions
of Ref.~\cite{ressat}. To complete this study we then quantify the impact of
the determined quark mass dependences of the mesons together with that of the
nucleon, which is already studied on the lattice, on the $NN$ observables via
an explict calculation of scattering lengths.  It is important to note that, to
our knowledge, no explicit
calculations for a dynamic generation of the omega-meson exist. 
Thus, we assume throughout that its quark mass dependence is the same as
of the $\rho$. Clearly, this should be refined in future studies.

The paper is structured as follows. In \Sec{sec:hadron_dep} we derive the
quark-mass dependence of nucleon and meson masses, which we use to calculate 
the impact of quark-mass variation on the two-nucleon potential in
\Sec{sec:NN_dep}. The theoretical uncertainties of our calculation are 
discussed in \Sec{sec:discuss}. From the two-nucleon observables we are 
able to calculate the quark-mass dependence of helium nuclei
(\Sec{sec:heavy_dep}). Finally, in Sections~\ref{sec:bbn} and \ref{sec:tau} we calculate the dependence
of primordial abundances on nuclear observables and combine this with the 
results of the previous sections to derive a limit on the variation of the light 
quark-mass at the time of big bang nucleosynthesis.

\section{Quark mass dependence of hadron masses}
\label{sec:hadron_dep}

Here, we study the quark mass dependence of the pertinent hadron masses.
The results for each hadron $H$ are most appropriately presented in terms of
the dimensionless parameters $K_H$ defined via
\begin{equation}
\frac{\delta M_H}{\delta m_f}=K_H^f\frac{M_H}{m_f} \ ,
\label{Kdef}
\end{equation}
evaluated at the physical point.
Here $m_f$ denotes the mass of the quark of flavor $f$
and $M_H$ denotes the mass of hadron $H$.
In what follows we will choose $f=q$ for the light quarks (in the isospin
limit) and $f=s$
for the strange quark.
 Note, although $m_f$
by itself is not renormalization group invariant, the
quantity of relevance here, namely $\delta m_f/m_f$ is,
for the same reason as quark mass ratios are well defined.

\subsection{Quark mass dependence of the nucleon mass}


Due to significant advances in lattice QCD the pion mass dependence of
especially the nucleon is now known to some precision. E.g. in
Ref.~\cite{latticemasses} the dependence of the nucleon mass on the pion mass
squared as calculated by the BMW collaboration is given. It is straightforward
 to extract from this the quantity $K_N^q$ (for the definition see
Eq.~(\ref{Kdef})). One finds $K_N^q = 0.04$. Note that more recent evaluations
from other lattice collaborations give similar results, as nicely reviewed
in \cite{AWL}. It is also pointed out in that reference that the nucleon mass 
can be well represented by a linear function of the pion mass in most lattice 
simulations, which is  at odds with the chiral constraints on this observable.

Alternatively one may determine $K_N^q$ from the pion--nucleon sigma
term, $\sigma_{\pi N}$. Ref.~\cite{chptmasses} finds
\begin{equation}\label{eq:sigmapiN}
\sigma_{\pi N}=44.9^{+1.8}_{-5.4} \ {\rm MeV} \ .
\end{equation}
On the other hand in Ref.~\cite{ollerverynew,alarcon} a value
\begin{equation}
\sigma_{\pi N}=59\pm 7 \ {\rm MeV} \ 
\end{equation}
is found. In what follows we use the first value as it is consistent with the
classical determination of Ref.~\cite{Gasser:1990ce} based on dispersion relations.
A completely reliable upgrade of the value from \cite{Gasser:1990ce} can only be 
obtained based on the recently proposed Roy-Steiner equations
for pion-nucleon scattering that allow for precise determination of the
pion-nucleon scattering amplitude in the physical region as well as
inside the Mandelstam triangle \cite{Ditsche:2012fv,Hoferichter:2012wf}. There is also
a large spread of values determined from lattice QCD which 
encompasses the range of values given above;
see Ref.~\cite{QCDSF} and the recent review by Kronfeld~\cite{Kronfeld:2012uk}.

Using the Feynman--Hellman theorem 
$\sigma_f = m_f \partial M_N/\partial m_f$,
one finds straightforwardly $K_N^f=\sigma_f/M_N$ and with that,
based on the numbers given above,
\begin{equation}
\label{Knucl}
K_N^q=0.048^{+0.002}_{-0.006} \ ,
\end{equation}
consistent with the number quoted above within 2$\sigma$.
The values given are significantly lower than those presented in
Refs.~\cite{craigs,craigs2} due to the unusually large value of
the $\pi N$ sigma term in those works.

\subsection{Quark mass dependence of meson masses}
\label{mesons}

Due to their nature as (pseudo)--Nambu-Goldstone bosons (NGBs) of the approximate chiral symmetry
of the strong interactions, the quark mass dependence of the members of the
pseudoscalar 
octet is given by standard ChPT, which is model-independent. At tree level, for the pion, one finds
\begin{equation}
K_\pi^q= 0.5\; ; \ K_\pi^s=0.
\label{pisimple}
\end{equation}
In our calculation, we have included higher order terms in the light (u,d)
quark mass dependence of the pion, using the $SU(2)$ ChPT expansion up to
two-loops~\cite{Bijnens:2006zp}. The strange quark mass dependence of the pion
and the quark mass dependence of the kaon and eta (which will be needed later
for the calculation of $K_\rho^f$ and $K_\sigma^f$) have been calculated using
$SU(3)$ ChPT to one loop~\cite{GL}. We remark, however, that these masses are
possibly affected by large higher order corrections -- this is an open issue
in three-flavor ChPT. For our study, we do not want to enter this
discussion but rather use the next-to-leading-order
(NLO) corrections based on the standard scenario
for the strange quark condensate.  To get at least some feeling for the
corresponding theoretical errors, we have estimated the size of the higher
order corrections by treating the ChPT expansions in two different ways: in
the first of them the expansions are written in terms of the physical masses
and decay constants of the NGB; in the second, they are written as a function
of the tree level constants $M_{0\pi}$, $M_{0K}$ and $F_0$. In our
calculations, these tree level constants are obtained by fitting the ChPT
expansions for the masses and decay constants of the NGB to their physical
values. Since the difference between the two treatments
corresponds to higher orders in the expansion, it serves as an estimate for the systematic error due to the truncation of the ChPT series. 

The values that we provide in the first line of \Tref{tab:Kestimates} for
$K_\pi^f$ are an average of the results obtained using $SU(2)$ and $SU(3)$,
the two methods of truncation mentioned above and different determinations for
the ChPT LECs~\cite{Amoros:2001cp,Colangelo:2001df,Bazavov:2009fk,Colangelo:2010et}. The
error is taken so that it covers all the results with their statistical
uncertainties, which arise from the errors of the LECs.  Following a similar
procedure we provide, also in \Tref{tab:Kestimates}, the $K^f_{F_\pi}$ values for the
pion decay constant $F_\pi$.

\begin{table}
  \centering
  \begin{tabular}{|c|c||c|c|}
  \hline
%
   $K_\pi^q$& $0.494\,^{+\,0.009}_{-\,0.013}$  &   $K_\pi^s$ & $0.00 \pm 0.05$ \\
   \hline
   $K_\rho^q$  & $0.058\pm0.002$  &   $K_\rho^s$   & $0.02\pm 0.04$ \\
  $K_\sigma^q$& $0.081\,\pm0.007$  &   $K_\sigma^s$ & $0.01 \pm 0.05$ \\ \hline
$K_{F_\pi}^q$ & $0.048 \pm 0.012$ & $K_{F_\pi}^s$ & $0.00\pm 0.06$  \\ 
\hline
  \end{tabular}
  \caption{Estimates for the $K_R^f$ coefficients and their uncertainties.}
\label{tab:Kestimates}
\end{table}

For the other light resonances the situation is less systematic, since
all of them are unstable and their correct description requires a pole in the complex
energy plane that cannot be obtained within the ChPT expansion, which, up to
some logarithmic terms, corresponds to an expansion in powers of the energy or
the meson masses. By construction, the amplitudes of ChPT are only
perturbatively unitary and valid only near threshold. In
Ref.~\cite{brunsmeissner} a formula to be used in chiral extrapolations for
vector meson masses was presented, however, the quality of lattice data was
not sufficient to pin down the slope of the quark mass dependence, which is
required here.  Of course, there are better data now
\cite{latticerhogood,latticerhoabove,lattice4}, so one could refresh the
analysis of Ref.~\cite{brunsmeissner}. Here, we follow another path, which can
also be used to explore the quark mass dependence of the rho and the
sigma. Employing dispersion relations for the inverse $\pi\pi$ scattering
amplitude and using ChPT to fix the subtraction constants --- where the
subtraction points can be chosen in a regime where ChPT is valid --- solves
both problems and, by generating poles, allows for the study of resonances
without a priori assumptions about their existence or nature. This is done in
a way consistent with the ChPT expansion without introducing spurious
parameters where an uncontrolled quark mass dependence may
appear\footnote{Clearly, the expansion is controlled only to the order of the
  chiral expansion used as input. Terms of higher order produced by the
  formalism are not necessarily correct~\cite{gassermeissner}, although, at
  least they will respect unitarity and the correct analytic structure of the
  amplitude.}. This method is called the inverse amplitude method (IAM)
\cite{IAM} and has been used to study both the $\sigma$ and the $\rho$ in an
SU(2) one loop treatment in Ref.~\cite{su2iam} and to two loops in
\cite{Pelaez:2010fj}. The SU(3) version of this study can be found in
Ref.~\cite{su3iam}.

Let us note that, within the IAM, all the dependence on the quark masses
appears through the NGB masses, which are explicitly present in the amplitudes,
both kinematically and in interaction vertices. Thus, we can calculate the
$K_R^f$ parameters for the resonances generated within the IAM by varying the
masses of the NGBs, whose dependence on the quark masses was discussed in
the previous section, and measuring the corresponding change on the position of the poles.

We have performed this calculation using $SU(2)$ and $SU(3)$ and different
sets of LECs obtained from IAM
fits~\cite{Nebreda:2011di,su2iam,Pelaez:2010fj,su3iam}. In each case, we have
changed the masses of the NGB using the two different methods for the
truncation of the ChPT expansion commented above. Our estimates for the
$K$-factors for $\rho$ and $\sigma$, given in the second part of
Table~\ref{tab:Kestimates}, are an average of the results combining these
different approaches, with errors taken to cover all the results. Let us note
that the description of the $\sigma$ depends more strongly on the chiral
loops, which are model-independent, and much less on the LECs. However, the
dependence of the $\rho$ resonance on the quark masses depends strongly on the
values of the LECs.  For that reason, for the central value of $K_\rho^q$ we
have only used the averaged results of the two best two-loop $SU(2)$ ChPT fits
in \cite{Pelaez:2010fj}, which we consider to be the most reliable, in
particular because they were fitted to three sets \cite{latticerhogood} of
lattice calculations of the $\rho$ mass dependence on the pion mass, which
were consistent among themselves\footnote{Let us nevertheless remark that
  there are other lattice calculations which are not quite compatible with
  these three because their $\rho$ masses fall systematically either above
  \cite{latticerhoabove} or below \cite{latticerhobelow} them. The ones
  falling below are somewhat harder to accommodate within the IAM, as
  explained in \cite{Pelaez:2010fj}}, and because the resulting values of the
LECs were more consistent with standard determinations and estimates. We refer
the reader to \cite{Pelaez:2010fj} for details. For the strange quark mass
dependence, we rely on the existing IAM one loop SU(3) calculations in
Ref.~\cite{su3iam}, but including the uncertainties as just described
above.

It should be stressed that the quark mass dependences we find are
significantly smaller than those given in Ref.~\cite{victoredward}.
In particular, in that reference a value of $K_\sigma^s=0.54$ is given
compared to our $-0.01$ (c.f.~\Tref{tab:Kestimates}). The origin
for this significant discrepancy is the assumption about the quark
structure of the $\sigma$ underlying the work of Ref.~\cite{victoredward}:
the $\sigma$ was assumed to be an SU(3) singlet. In our case on the
other hand the $\sigma$ emerges from non--perturbative $\pi\pi$ interactions,
which give only a very small dependence on the strange quark mass.

As becomes clear from \Tref{tab:Kestimates}, for all relevant quantities
the variation with respect to the strange quark mass is smaller that the 
corresponding uncertainty. In addition, some quark mass variations driven by an
external scale will lead to a relative change in the strange sector suppressed
additionally by a factor $m_q/m_s\sim 1/25$. In what follows we will therefore
only study the effect of a  variation of the light quark masses on the $NN$ 
potential.

\section{Impact on the variation of the two-nucleon potential}
\label{sec:NN_dep}

The changes in the meson masses cannot be directly connected to
their impact on BBN. The quantity of relevance is the resulting variation of the
two-nucleon ($NN$)
interaction and, especially, its impact on nuclear binding energies. However,
the connection of meson masses to the $NN$ interaction is not clear cut. In
Ref.~\cite{victorandbob} the phenomenological V18 interaction was used as a
basis, where the insights of Refs.~\cite{craigs,craigs2} were translated into
a variation of the model parameters. Although it provides some understanding
of the sensitivity of the $NN$ interaction on the variation of quark mass
parameters, such a calculation is neither systematic nor complete.
On the other hand, in Ref.~\cite{platterbedaque} an EFT approach is chosen;
however, there the input of the quark mass dependence of $NN$ scattering
lengths is taken from other sources. Our goal is to improve our understanding of
the
quark mass dependence of the $NN$ observables using as input the $K$-factors
given in \Tref{tab:Kestimates}.

As outlined in the introduction, we do not do the full dispersion theoretical 
treatment of Refs.~\cite{john1,john2}, but make connection to the $NN$ force
via the method of resonance saturation: in Ref.~\cite{ressat} explicit
expressions are given that allow one to express the values of the $NN$
contact terms in terms of meson masses and coupling constants. Thus,
the quark mass dependences given above can be implemented straightforwardly.

\subsection{Quark mass dependence of the pion exchange contributions}
\label{sec:ga}

The long-range part of the $NN$ potential up to next-to-next-to-leading
order (N$^2$LO) in the chiral expansion is driven by the exchange
of one and two pions. In the exact isospin limit, the one-pion ($1\pi$) exchange
potential at N$^2$LO has the form
\beq
\label{VLO}
V_{\rm 1 \pi} = -\frac{1}{4 F_\pi^2} \left( g_A - 2 d_{18} M_\pi^2
\right)^2  \;\fet \tau_1 \cdot \fet \tau_2  \;
\frac{(\vec
  \sigma_1 \cdot \vec q)( \vec \sigma_2 \cdot \vec q)}{\vec q {\,}^2 +M_\pi^2}  \,,
\eeq
where $\sigma_i$
denote the Pauli spin matrices, $\vec q
= \vec p \, ' - \vec p$ is the nucleon momentum transfer and $\vec{p}$
($\vec{p}~'$) refers to initial (final) nucleon momenta in the center-of-mass system. Further,
$F_\pi$ and $g_A$ denote the pion decay and the nucleon
axial coupling constants, respectively, while $d_{18}$ is a low-energy
constant from $\mathcal{L}_{\pi N}^{(3)}$ that controls the leading
contribution to Goldberger-Treiman discrepancy.  
Employing the spectral function regularization
as detailed in Ref.~\cite{Epelbaum:2003gr}, the non-polynomial part of
the two-pion ($2\pi$) exchange potential has the form 
\beqa
\label{VNLO}
V^{(2)}_{\rm 2N} &=&  - \frac{ \fet{\tau}_1 \cdot \fet{\tau}_2 }{384 \pi^2 F_\pi^4}\,
L^{\tilde \Lambda} (q) \, \left( 4M_\pi^2 (5g_A^4 - 4g_A^2 -1)
+ \vec{q}\, ^2 (23g_A^4 - 10g_A^2 -1)
+ \frac{48 g_A^4 M_\pi^4}{4 M_\pi^2 + \vec{q} \, ^2} \right) \nn
&-&  \frac{3 g_A^4}{64 \pi^2 F_\pi^4} \,L^{\tilde \Lambda} (q)  \, 
\left(  \vec \sigma_1 \cdot \vec q  \, \vec \sigma_2 \cdot \vec q   -  \vec \sigma_1 \cdot\vec
\sigma_2  \, \vec{q} \, ^2 \right) \,, \nn 
V^{(3)}_{\rm 2N} &=&  -\frac{3g_A^2}{16\pi F_\pi^4}  \big( 2M_\pi^2(2c_1 -c_3)
-c_3 \vec{q} \, ^2 \big) 
 (2M_\pi^2+\vec{q} \, ^2)
 A^{\tilde \Lambda} (q)   
- \frac{g_A^2 c_4 }{32\pi F_\pi^4} \, \fet{ \tau}_1 \cdot \fet{
  \tau}_2 \,  (4M_\pi^2 + q^2) \nn
&\times & A^{\tilde \Lambda}(q)\, 
\big( \vec \sigma_1 \cdot \vec q\, \vec \sigma_2 \cdot \vec q\, 
-\vec{q} \, ^2 \, \vec \sigma_1 \cdot\vec \sigma_2 \big)\,,
 \eeqa
with the loop functions $L^{\tilde \Lambda} (q)$ and $A^{\tilde \Lambda} (q)$ 
defined as
\beqa
\label{def_LA}
L^{\tilde \Lambda} (q) &=& \theta (\tilde \Lambda - 2 M_\pi ) \,
\frac{\omega}{2 q} \, \ln \frac{\tilde \Lambda^2 \omega^2 + q^2 s^2 + 2 \tilde
  \Lambda q \omega s}{4 M_\pi^2 ( \tilde \Lambda^2 + q^2)} \,, \quad \omega =
\sqrt{4 M_\pi^2 + \vec{q}\, ^2}\,, \quad s = \sqrt{\tilde \Lambda^2 - 4
  M_\pi^2}\,, \nn A^{\tilde \Lambda} (q) &=& \theta (\tilde \Lambda - 2 M_\pi
) \, \frac{1}{2 q} \, \arctan \frac{q ( \tilde \Lambda - 2 M_\pi )}{q^2 + 2
  \tilde \Lambda M_\pi}\,.  \eeqa 
Here and in what follows, the $c_i$ are the LECs
from the order-$Q^2$ (dimension two) pion-nucleon Lagrangian and $\tilde
\Lambda$ denotes the cutoff in the spectral representation, see
Ref.~\cite{Epelbaum:2003gr}.  In addition to the explicit $M_\pi$-dependence,
at N$^2$LO one also needs to take into account the implicit one resulting from
the chiral expansion of $g_A$ and $F_\pi$ in the $1\pi$-exchange potential in
Eq.~(\ref{VLO}). We use the NLO result for the chiral expansion of the pion
decay constant as it is appropriate at the order we are working: \beq F_\pi =
F \left( 1 + \frac{M_\pi^2}{(16 \pi^2 F^2)} \bar l_4 + \mathcal{O} \left(
M_\pi^4 \right) \right)\,, \eeq where $F$ denotes the pion decay constant in
the chiral limit.  For the LEC $\bar l_4$ we adopt the value $\bar l_4 = 4.3$
from Ref.~\cite{Gasser:1983yg}. Using
$F_\pi = 92.2$ MeV, this leads to the chiral-limit value $F=86.2$
MeV. Notice that the resulting NLO value for the $K$-factor $K^q_{F
  \pi} = 0.065$  is slightly larger than the one given in
\Tref{tab:Kestimates}.   

\begin{figure}[t!]
\begin{center}
\epsfig{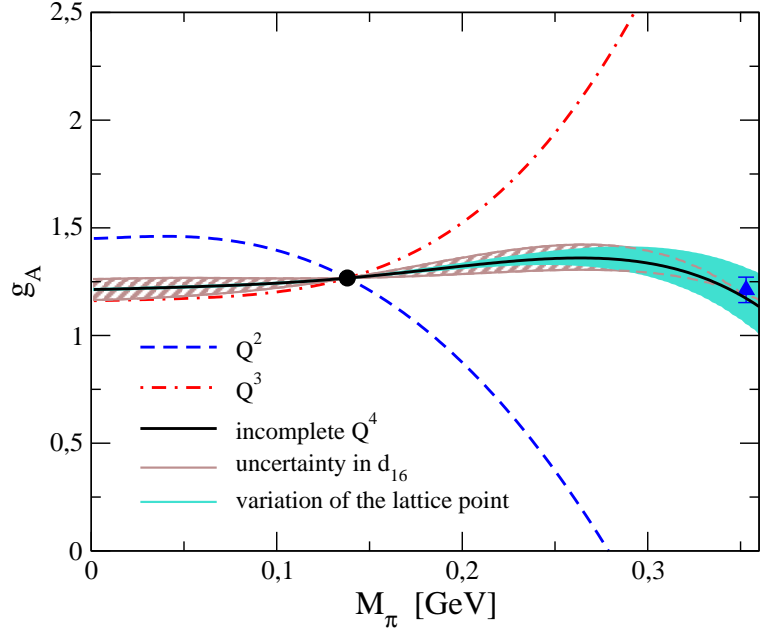}
\caption{Quark mass dependence of $g_A$ at various orders in the
  chiral expansion. Also shown is the lowest mass lattice point from
  Ref.~\cite{Edwards:2005ym}.  The hatched band corresponds  to the variation of $\bar d_{16}$ in
 the range $\bar d_{16}= -0.91$ to $-2.61$~GeV$^{-2}$, see
 Ref.~\cite{Fettes:2000fd}. The light shaded band results from a $10\%$ variation
 of the lattice point used to fix the LEC of the $M_\pi^4$ operator in
 Eq.~(\ref{ga}).}\label{fig:ga}
\end{center}
\end{figure}

Contrary to  the pion decay constant and the
nucleon mass, the chiral expansion for $g_A$ is known to 
converge rather slowly, see Fig.~\ref{fig:ga}. 
In particular,
taking into account the leading (i.e.~order $\mathcal{O} (M_\pi^2)$) correction to
its value at the chiral limit and
adopting the value for the low-energy constant $\bar d_{16} =-1.76$
GeV$^{-2}$ obtained from the reaction $\pi N \to  \pi \pi N$  \cite{Fettes:2000fd} leads to a
very strong quark-mass dependence of $g_A$ near the physical point. On
the other hand, lattice QCD calculations indicate that the behavior
of $g_A$ with $M_\pi$ is rather flat. As discussed in Ref.~\cite{Bernard:2006te},
such a flat behavior of $g_A$, consistent with the lowest-mass lattice
data point from Ref.~\cite{Edwards:2005ym} corresponding to $M_\pi
=353$ MeV 
\footnote{We emphasize, however, that the convergence at such
pion masses is problematic, as discussed in detail in the review
\cite{Bernard:2007zu}. } 
can, in principle, be achieved at the two-loop
level. In order to provide an accurate representation of the quark mass
dependence of the $1\pi$-exchange potential, we use in the present
study the complete order-$Q^3$ result   
accompanied by one order-$Q^4$ contact term,  whose strength is adjusted to reproduce the
lowest-mass lattice data from Ref.~\cite{Bernard:2006te}:
\beqa
\label{ga}
g_A &=& g_0 \left[ 1 + \left( \frac{4}{g_0} \bar d_{16} - \frac{1}{(4 \pi
    F)^2} \left( g_0^2 + \left( 2 + 4 g_0^2 \right) \ln
    \frac{M_\pi}{\bar M_\pi} \right) \right) M_\pi^2 \right. \nn
&&{} + \left. \frac{1}{24 \pi F^2 m_0} \left( 3 + 3 g_0^2 - 4 m_0 c_3 + 8 m_0 c_4
\right) M_\pi^3 + \beta M_\pi^4
\right]\,.
\eeqa
Here, $g_0$, $F$ and $m_0$ refer to the chiral-limit values of the
nucleon axial vector coupling, pion decay coupling and the nucleon
mass, respectively. We use the same values of the LECs $c_i$ as in 
Ref.~\cite{Bernard:2006te} namely $c_3 = -4.7$ GeV$^{-1}$, $c_4 = 3.5$
GeV$^{-1}$. \footnote{Notice that slightly different values of these
  LECs are adopted in the chiral $NN$ potential. We have verified that
  using these different values for $c_{3,4}$ and re-adjusting the
  parameter $\beta$ leads to a very similar $M_\pi$-dependence of
  $g_A$. The induced difference in the two-nucleon observables is
  significantly beyond the theoretical uncertainty of our analysis.} 
Further, $\bar M_\pi = 138$ MeV is the
physical value of the pion mass in the isospin limit while $\beta$ is a
linear combination of the order-$Q^4$ LECs.  We emphasize that
the above expression does not correspond to the full order-$Q^4$
result since we do not include the order-$Q^4$ chiral logarithms. We
have verified numerically that the effect of these logarithms is 
largely compensated by the $\beta$-term when the later is tuned to
reproduce the lattice point. Indeed, one observes that the solid line
in Fig.~\ref{fig:ga} corresponding to the pion mass dependence of $g_A$
adopted in the present work is very similar to the more complete
calculations of  Ref.~\cite{Bernard:2006te} shown in Fig.~2 of that work.   
Further, Fig.~\ref{fig:ga} also shows the uncertainty associated with
the variation of $\bar d_{16}$ in the range  $\bar d_{16}= -0.91$ to $-2.61$~GeV$^{-2}$ 
\cite{Fettes:2000fd} and the variation of the lattice point by
 $10\%$. 

The value of the nucleon mass in the
chiral limit can be obtained from the order-$Q^3$ expression
\beq
\label{mnucl}
m = m_0 + 4 c_1 M_\pi^2 + \frac{3 g_A^2}{32 \pi F_\pi^2} M_\pi^3 +
\mathcal{O} \left( M_\pi^4 \right)\,.
\eeq   
Using $g_A=1.267$ and  $c_1 = -0.81$ GeV$^{-1}$ leads to 
$m_0=892$ MeV. Note that the value of $c_1$ used here
is consistent with the small sigma-term, $\sigma_{\pi N}=45\,$MeV,
cf. Eq.~(\ref{eq:sigmapiN}). We further emphasize that the
resulting $K$-factor, $K_N^q=0.042$, is consistent (within the error bars) with the value
given in  Eq.~(\ref{Knucl}).

\subsection{Quark mass dependence of the short-range terms}

The short-range potential in the $^1S_0$ and $^3S_1- ^3D_1$ channels
up to N$^2$LO has the form
\beqa
V_{1S0}^{\rm short} &=& \tilde C_{1S0} + C_{1S0} (p^2 + {p '} ^2) \,, \nn
V_{3S1}^{\rm short} &=& \tilde C_{3S1} + C_{3S1} (p^2 + {p '} ^2) \,, \nn
V_{\epsilon 1}^{\rm short} &=& C_{\epsilon 1} \, p^2 \,,
\eeqa
where $p \equiv | \vec p |$,  $p '\equiv | \vec p  ' |$ refer to the
in-coming and out-going momenta in the center-of-mass system and 
$\tilde C_i$, $C_i$ are $M_\pi$-dependent coefficients\footnote{Of course, the
LECs do not depend on the quark masses. The coefficients used here subsume the
coefficients of the LO four-nucleon operators plus their first pion mass
dependent corrections that are generated by the same operators times $M_\pi^2$.}.   
The quark mass dependence of these operators can,
in principle, be calculated straightforwardly in chiral
perturbation theory \cite{chiralextrapol_bonn}. The problem is,
however, that the coefficients in front of the contact
operators  $\propto M_\pi^2$ are unknown. In
\cite{chiralextrapol_bonn}, the corresponding LECs were estimated by
means of na\"ive dimensional analysis which, however, resulted in a very
large theoretical uncertainty for two-nucleon observables. 
In order to avoid
this difficulty, we follow here a completely different approach
and refrain from doing an explicit chiral expansion for contact
operators. Instead, we make use of resonance saturation of contact 
interactions~\cite{ressat} and employ the $M_\pi$-dependence for
the masses of the heavy mesons discussed
in Sec.~\ref{mesons}.   

Resonance saturation for contact $NN$ operators up to N$^2$LO is
discussed in detail in Ref.~\cite{ressat}. In that work strongly
reduced values of the LECs $c_i$ were adopted in order to circumvent
a very strong attraction resulting from the isoscalar central two-pion
($2\pi$) exchange potential calculated using dimensional regularization. As
discussed in Ref.~\cite{Epelbaum:2003gr}, the strong attraction in the
$2\pi$-exchange potential at intermediate and shorter distances can be
traced back to the large-mass components in the spectrum which cannot
be described reliably within the framework of chiral EFT. In the
chiral potentials of Refs.~\cite{Epelbaum:2003xx,Epelbaum:2004fk}, the
unphysical high-mass components
in the two-pion exchange spectrum are cut off by means of the spectral
function regularization. We now repeat the analysis of
Ref.~\cite{ressat} for the actual version of the chiral N$^2$LO
potentials and test the validity and accuracy of the resonance
saturation hypothesis.  

Here and in what follows, we consider the Bonn-B \cite{bonnpot} potential as
a typical representative of one-boson-exchange (OBE) models. Its
long-range part is given by $1\pi$-exchange whereas shorter-distance
physics is expressed as a sum over contributions from the exchange
of heavier mesons. For nucleon momentum transfers well below the masses of the
exchanged mesons, one can interpret such exchange diagrams in terms of
local contact operators with an increasing number of derivatives
(momentum insertions). The LECs accompanying the resulting contact
interactions are then expressed in terms of the meson masses,
meson-nucleon coupling constants and the corresponding form-factors. In
order to allow for a meaningful comparison between the chiral and OBE
potentials, one needs to properly account for the
chiral $2\pi$-exchange potential, which contributes to the chiral
potential but is absent in the OBE models. Here we follow the strategy
of Ref.~\cite{ressat}  and expand the $2\pi$-exchange potential of
Eq.~(\ref{VNLO}) in
powers of momenta. This induced contributions to the LECs entering 
the $^1S_0$ and $^3S_1 - ^3D_1$ partial waves read:
\beqa
\delta \tilde C_{1S0}^{(2)} &=&\frac{\left(-8 g_A^4+4 g_A^2+1\right)
  M_\pi^2 \sqrt{
\tilde \Lambda ^2-4 M_\pi^2}}{24 \pi  F_\pi^4 \tilde \Lambda
   }\,, \nn
\delta  C_{1S0} ^{(2)} &=&\frac{\sqrt{\tilde \Lambda ^2-4 M_\pi^2} \left(\left(-88 g_A^4+17 g_A^2+2\right)
   \tilde \Lambda ^2+2 \left(8 g_A^4-4 g_A^2-1\right) M_\pi^2\right)}{144 \pi 
   F_\pi^4 \tilde \Lambda ^3} \,, \nn
\delta \tilde C_{3S1} ^{(2)} &=& \frac{\left(8 g_A^4-4 g_A^2-1\right) M_\pi^2
  \sqrt{\tilde \Lambda ^2-4
   M_\pi^2}}{8 \pi  F_\pi^4 \tilde \Lambda }\,, \nn
\delta  C_{3S1} ^{(2)} &=&\frac{\sqrt{\tilde \Lambda ^2-4 M_\pi^2} \left(\left(40 g_A^4-17 g_A^2-2\right)
   \tilde \Lambda ^2+\left(-16 g_A^4+8 g_A^2+2\right) M_\pi^2\right)}{48 \pi 
   F_\pi^4 \tilde \Lambda ^3}\,, \nn
\delta  C_{\epsilon 1}^{(2)}  &=&-\frac{g_A^4 \sqrt{\tilde \Lambda ^2-4 M_\pi^2}}{4
  \sqrt{2} \pi  F_\pi^4 \tilde \Lambda 
   } \, ,
\eeqa
at NLO and 
\beqa
\delta \tilde C_{1S0} ^{(3)} &=&   \frac{3 g_A^2 M_\pi^3 (2 c_1-c_3 ) (2
  M_\pi-\tilde \Lambda )}{4 F_\pi^4
  \tilde  \Lambda }  \,, \nn
\delta  C_{1S0} ^{(3)} &=& -\frac{g_A^2 M_\pi (2 M_\pi-\tilde \Lambda ) \left(\tilde \Lambda ^2
    (-10 c_1 +11
   c_3 -4 c_4)+4 M_\pi^2 (2 c_1-c_3)+2 \tilde \Lambda  M_\pi (2
   c_1-c_3)\right)}{16 F_\pi^4 \tilde \Lambda ^3}      \,, \nn
\delta \tilde C_{3S1} ^{(3)} &=&  \frac{3 g_A^2 M_\pi^3 (2 c_1-c_3) (2 M_\pi-\tilde \Lambda )}{4 F_\pi^4
   \tilde \Lambda }  \,, \nn
\delta  C_{3S1} ^{(3)} &=& -\frac{g_A^2 M_\pi (2 M_\pi-\tilde \Lambda ) \left(\tilde \Lambda ^2 (-10 c_1+11
   c_3-4 c_4)+4 M_\pi^2 (2 c_1-c_3)+2 \tilde \Lambda  M_\pi (2
   c_1-c_3)\right)}{16 F_\pi^4 \tilde \Lambda ^3}  \,, \nn
\delta  C_{\epsilon 1} ^{(3)} &=& \frac{c_4  g_A^2 M_\pi (\tilde \Lambda -2 M_\pi)}{2 \sqrt{2} F_\pi^4 \tilde \Lambda }\,,
\eeqa
at N$^2$LO. In the limit $\tilde \Lambda \to \infty$ corresponding to
dimensional regularization, the 
above expressions agree with the ones given in Ref.~\cite{ressat}. The
size of the $2\pi$-exchange-induced contributions to the LECs for the
two extreme values of the spectral function cutoff $\tilde \Lambda$
can be read off \Tref{tab1}. 
\begin{table*}[t] 
\begin{center}
\begin{tabular*}{1.0\textwidth}{@{\extracolsep{\fill}}|l|cc|cc|}
    \hline
 LEC   & $Q^2$, $\tilde \Lambda = 500$ MeV    &  $Q^2$, $\tilde
 \Lambda = 700$ MeV  &  $Q^3$, $\tilde \Lambda = 500$ MeV  &  $Q^3$, $\tilde \Lambda = 700$ MeV  \\[1ex]
\hline 
$\delta \tilde{C}_{1S0}$  &  $-$0.004 & $-$0.005 & $-$0.004 & $-$0.005 \\ 
$\delta {C}_{1S0}$         & $-$0.534 & $-$0.592 & $-$0.365  &  $-$0.500\\ 
$\delta \tilde{C}_{3S1}$  & 0.013 & 0.014 & $-$0.004 & $-$0.005 \\ 
$\delta {C}_{3S1}$      & 0.594 & 0.663 &  $-$0.365 & $-$0.500 \\ 
$\delta {C}_{\epsilon 1}$ & $-$0.178  & $-$0.196 & 0.170 & 0.229 \\ 
\hline  
  \end{tabular*}
\caption{Contributions to the LECs $\tilde C_i$ and $C_i$ induced by
  the NLO and N$^2$LO $2\pi$-exchange potential. The $\tilde C_i$ are in 10$^4$ GeV$^{-2}$ 
and the $C_i$ in 10$^4$ GeV$^{-4}$.
}\label{tab1}
\end{center}
\end{table*}

After these preparations, we are now in the position to test the
resonance saturation hypothesis for $\tilde C_i$ and $C_i$. The
contributions of the various mesons to the contact operators can be
obtained by carrying out partial wave decomposition of the expressions
for the boson exchange contributions given in Ref.~\cite{ressat} and
expanding the results in powers of momenta.  In Tables~
\ref{tab2} and \ref{tab3}, the $2\pi$-exchange-corrected values of these LECs for
five cutoff combinations 
\beqa
\label{cutoffs_3N}
\mbox{NLO}: &&  \{ \Lambda, \; \tilde \Lambda \} = \{ 400, \; 500 \},  \; \{ 550, \; 500 \},  \; 
\{ 550, \; 600 \},  \;\{ 400, \; 700 \},  \; \{ 550, \; 700 \}\,, \nn
\mbox{N$^2$LO}: &&  \{ \Lambda, \; \tilde \Lambda \} = \{ 450, \; 500 \},  \; \{ 600, \; 500 \},  \; 
\{ 550, \; 600 \}, \; \{ 450, \; 700 \},  \; \{ 600, \; 700 \}\,, \nn
&&
\eeqa
 are listed together with the values resulting from resonance saturation based on the
 Bonn-B model. 
\begin{table*}[t] 
\begin{center}
\begin{tabular*}{1.0\textwidth}{@{\extracolsep{\fill}}|l|ccccc|c|}
    \hline
    & fit 1  & fit 2 & fit 3 & fit 4 & fit 5 & Bonn-B  \\[1ex]
\hline 
$\tilde{C}_{1S0} + \delta \tilde{C}_{1S0}^{(2)}$  &  $-$0.161  & $-$0.066  & $-$0.095  & $-$0.161  &
$-$0.111 & $-$0.117 \\ 
${C}_{1S0}+ \delta {C}_{1S0}^{(2)} $         &  0.974&  1.574 & 1.457 & 1.008 & 1.386 & 1.276\\ 
$\tilde{C}_{3S1}+ \tilde{C}_{3S1}^{(2)} $  & $-$0.169  & $-$0.136 & $-$0.135 & $-$0.167 & $-$0.135 & $-$0.101\\ 
${C}_{3S1}+ \delta {C}_{3S1}^{(2)} $      & 0.356  & 0.256 & 0.231 & 0.280 & 0.221 & 0.660 \\ 
${C}_{\epsilon 1}+ \delta {C}_{\epsilon 1}^{(2)} $ & $-$0.390  & $-$0.332 & $-$0.325 & $-$0.373 & $-$0.321 & $-$0.410\\ 
\hline  
  \end{tabular*}
\caption{LECs $\tilde C_i$ and $C_i$ from the NLO chiral potential
  for different cutoff combinations (fits 1 to 5 as defined
  in Eq.~(\ref{cutoffs_3N})) 
  corrected by the induced contributions of the $2\pi$-exchange
  potential.  Also shown are values resulting from resonance
  saturation using the Bonn-B model (last column).
 The $\tilde C_i$ are in 10$^4$ GeV$^{-2}$ 
and the $C_i$ in 10$^4$ GeV$^{-4}$.
}\label{tab2}
\end{center}
\end{table*}
\begin{table*}[t] 
\begin{center}
\begin{tabular*}{1.0\textwidth}{@{\extracolsep{\fill}}|l|ccccc|c|}
    \hline
    & fit 1  & fit 2 & fit 3 & fit 4 & fit 5 & Bonn-B  \\[1ex]
\hline 
$\tilde{C}_{1S0}+ \delta \tilde{C}_{1S0}^{(2)} + \delta \tilde{C}_{1S0}^{(3)}$  &  $-$0.161  & $-$0.116  & $-$0.159  & $-$0.163  &
$-$0.161 & $-$0.117 \\ 
${C}_{1S0}+ \delta {C}_{1S0}^{(2)} + \delta {C}_{1S0}^{(3)}$         &  1.164&  1.368 & 1.243 & 1.321 & 1.321 & 1.276\\ 
$\tilde{C}_{3S1}+ \delta \tilde{C}_{3S1}^{(2)} + \delta \tilde{C}_{3S1}^{(3)}$  & $-$0.162  & $-$0.127 & $-$0.137 & $-$0.164 & $-$0.130 & $-$0.101\\ 
${C}_{3S1}+ \delta {C}_{3S1}^{(2)} + \delta {C}_{3S1}^{(3)}$      & 0.574  & 0.423 & 0.523 & 0.720 & 0.568 & 0.660 \\ 
${C}_{\epsilon 1}+ \delta {C}_{\epsilon 1}^{(2)} +
\delta {C}_{\epsilon 1}^{(3)}$ & $-$0.425  & $-$0.363 & $-$0.395 & $-$0.467 & $-$0.409 & $-$0.410\\ 
\hline  
  \end{tabular*}
\caption{LECs $\tilde C_i$ and $C_i$ from the N$^2$LO chiral potential
  for different cutoff combinations (fits 1 to 5 as defined
  in Eq.~(\ref{cutoffs_3N})) 
  corrected by the induced contributions of the $2\pi$-exchange
  potential.  Also shown are values resulting from resonance
  saturation using the Bonn-B model (last column).
 The $\tilde C_i$ are in 10$^4$ GeV$^{-2}$ 
and the $C_i$ in 10$^4$ GeV$^{-4}$.
}\label{tab3}
\end{center}
\end{table*}
For the considered cutoff combinations, all LECs are reasonably
well described in terms of resonance saturation. The agreement is
better at N$^2$LO, most notably for the LEC $C_{3S1}$. The
estimations based on  resonance saturation yield the numbers which are
typically within 20 -- 30\% of the true values except for the LO LECs $\tilde C_{i}$
which appear to be somewhat more strongly underestimated.   For the sake of
completeness, we also list in Table~\ref{tab4} the contributions from
individual mesons exchanges in the Bonn-B model, see also
Ref.~\cite{ressat}. 
\begin{table}[htb]
\begin{center}
\begin{tabular*}{1.0\textwidth}{@{\extracolsep{\fill}}|l|ccccc|c|}
\hline
LEC  &  $\eta$  & $\sigma$ & $\delta$ & $\omega$ & $\rho$ & sum  \\
    \hline
$\tilde{C}_{1S0}^{\rm res}$  &  $0.000$  &  $-0.392$
             &  $-0.023$ & $0.287$  & $0.011$ & $-0.117$ \\
${C}_{1S0}^{\rm res}$        &  $0.033$  &  $1.513$
             &  $0.036$ & $-0.560$  & $0.254$ & $1.276$ \\
$\tilde{C}_{3S1}^{\rm res}$  & $0.000$  &  $-0.424$
             &  $0.070$ & $0.287$  & $-0.034$ & $-0.101$ \\
${C}_{3S1}^{\rm res}$        &  $-0.011$  &  $1.030$
             &  $-0.108$ & $-0.777$  & $0.526$ & $0.660$ \\
${C}_{\epsilon 1}^{\rm res}$    &  $-0.032$  &  $0.000$
             &  $0.000$ & $0.077$  & $-0.455$ & $-0.410$ \\
\hline
  \end{tabular*}
\vspace{0.3cm}
\caption{Contributions of the various boson exchanges to the LECs
for the Bonn-B potential and the corresponding sum.
The $\tilde{C}_i$ are in $10^{4}~$GeV$^{-2}$ and the $C_i$ in
$10^{4}~$GeV$^{-4}$.
}
\label{tab4}
\end{center}
\end{table}

\begin{figure}[t!]
\begin{center}
\epsfig{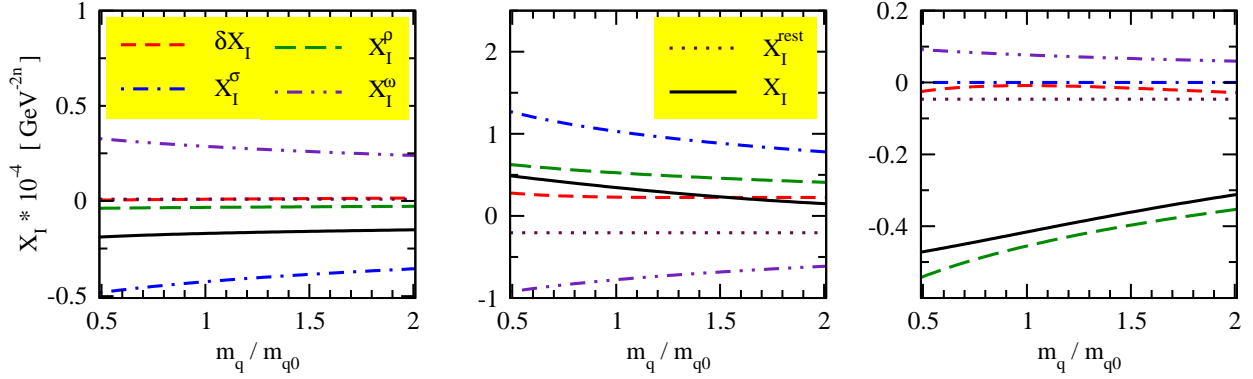}
\caption{Various contributions to the quark mass dependence of the LECs $\tilde C_{3S1}$ (left
  panel), $C_{3S1}$ (middle panel) and $C_{\epsilon 1}$ (right panel)
  for fit 1 at N$^2$LO as discussed in the text. Here, $m_{q0}$ denotes the
  physical value of the light quark mass.
  }\label{fig:LECrun}
\end{center}
\end{figure}

The observed reasonably good representation of the LECs accompanying the
short-range operators  in terms of heavy-meson exchanges justifies
modelling the chiral extrapolations for  the corresponding
operators in terms of quark/pion mass dependence of the heavy mesons
as discussed in \Sec{mesons}. More precisely, we assume that
the resonance saturation hypothesis remains  valid at unphysical values of
the quark/pion masses, that is 
\beq
\label{run_LEC}
X_I  (M_\pi) + \delta X_I ( M_\pi ) = X_I^\sigma  (M_\pi) +
X_I^\rho  (M_\pi) +  X_I^\omega  (M_\pi) + X_I^{\rm rest}\,.
\eeq
Here $X$ stays for $C$, $\tilde C$ and $I= \{ 1S0, \; 3S1, \;
\epsilon 1 \}$ and $\delta X_I ( M_\pi ) = \delta X_I^{(2)} ( M_\pi )
$   at NLO and  $\delta X_I ( M_\pi ) = \delta X_I^{(2)} ( M_\pi ) + \delta X_I^{(3)} ( M_\pi )
$ at N$^2$LO. For the resonance contributions, we take into
account the quark mass dependence of the $\sigma$- and $\rho$-meson
masses as given in \Tref{tab:Kestimates} and assume $K_\omega^q =
K_\rho^q$ for the $\omega$-meson~\cite{craigs2,craigs}. 
Neglecting the quark mass dependence of the $\eta$-
and $\delta$-mesons is justified by their small contributions to the
LECs. The resulting error is expected to be well below the theoretical
uncertainty of our analysis. Notice that in  $X_i^{\sigma, \rho, \omega}$, we also
take into account the quark mass dependence of the nucleon mass. The
last term on the right-hand side of Eq.~(\ref{run_LEC}), $X_i^{\rm rest}$, denotes the
contributions to the LECs not related to the heavy-boson
exchanges. The (unknown) $M_\pi$-dependence of $X_I^{\rm rest}$  is
neglected in the present work. This, of course, only makes sense if
$X_I^{\rm rest}$ is small
compared to the $X_I$, i.e.~if the LECs are well described in terms of 
resonance saturation. This is indeed the case for both the NLO and
N$^2$LO potentials.  As a representative example, we show in
Fig.~\ref{fig:LECrun} the individual contributions to the quark mass dependence
of $\tilde C_{3S1}$, $C_{3S1}$ and $C_{\epsilon_1}$ resulting from
Eq.~(\ref{run_LEC}) for fit 1 at N$^2$LO. 
While strong cancellations between the $\sigma$- and
$\omega$-contributions  are observed for $\tilde C_{3S1}$ and
$C_{3S1}$, the LEC $C_{\epsilon 1}$ is largely saturated by the
$\rho$-meson. Notice that as a result of the cancellations, 
there is a sizeable uncertainty in the $m_q$-dependence of 
$C_{3S1}$ associated with the non-resonance contribution of the last
term in Eq.~(\ref{run_LEC}).

\subsection{Quark mass dependence of the S-wave $NN$ observables}

Having specified the quark mass dependence of the $NN$ potential, we now
turn to the chiral extrapolations of two-nucleon S-wave observables. We
calculate the $NN$ phase shifts and mixing angles by solving the 
nonrelativistic\footnote{Relativistic corrections to the two-nucleon
  Green's function need to be taken into account starting
  from N$^3$LO which is beyond the scope of the present work.} 
Lippmann-Schwinger (LS) equation in the partial wave basis
\beq\label{LSeq}
T^{sj}_{l'l} (p',p) = V^{sj}_{l'l} (p',p) +  \sum_{l''} \,
\int_0^\infty \frac{dp'' \, {p''}^2}{(2 \pi )^3} \,  V^{sj}_{l'l''} (p',p'')
\frac{m}{p^2-{p''}^2 +i\eta} T^{sj}_{l''l} (p'',p)~,
\eeq
as $\eta \to 0^+$.
The relation between the on-shell $S$- and $T$-matrices is given by 
\be
S_{l' l}^{s j} (p) = \delta_{l' l} - \frac{i}{8 \pi^2} 
\, p \, m \,  T_{l' l}^{s j} (p)~.
\ee
Thus, the quark-/pion-mass dependence in the observables emerges from 
both the nucleon mass entering the $NN$ Green's  function and the
potential. 

\begin{figure}[t!]
\begin{center}
\epsfig{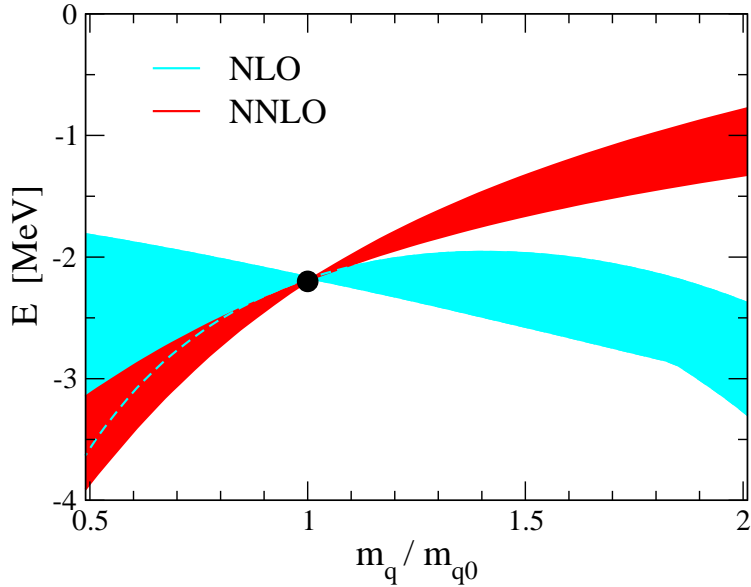}
\caption{Quark mass dependence of the deuteron binding energy
  at NLO (light-shaded band) and N$^2$LO
  (dark-shaded band). The bands correspond to the cutoff variation as
  discussed in the text.  The solid dot shows the deuteron binding
  energy at the physical quark mass. 
  }\label{fig:BErun}
\end{center}
\end{figure}

\begin{figure}[t!]
\begin{center}
\epsfig{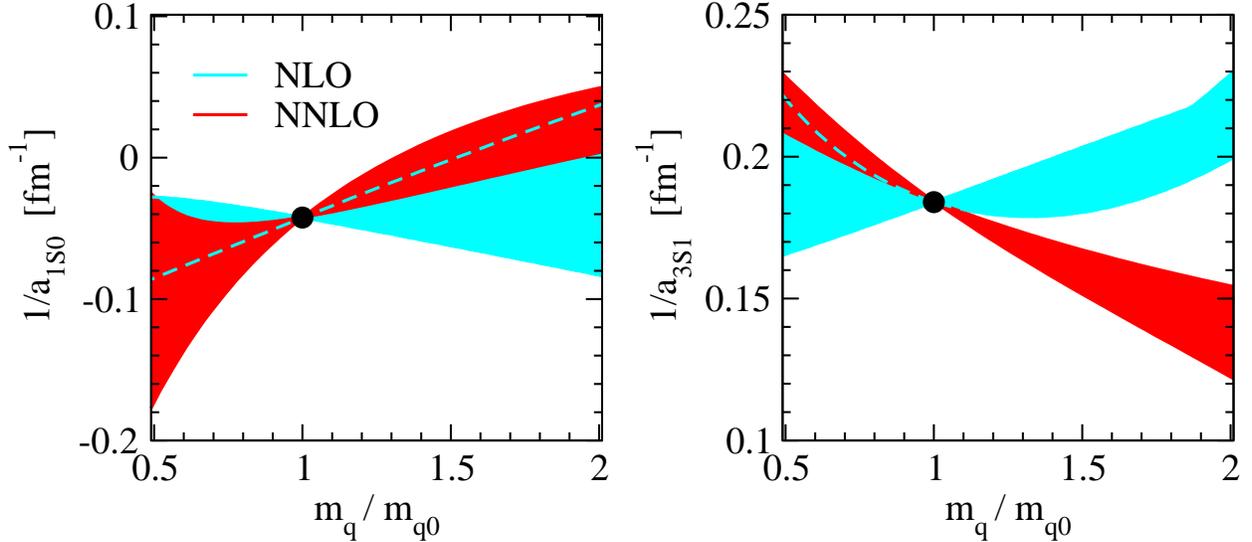}
\caption{Quark mass dependence of the inverse S-wave scattering
  lengths at NLO (light-shaded band) and N$^2$LO
  (dark-shaded band). The bands correspond to the cutoff variation as
  discussed in the text.  The solid dots show the inverse scattering
  lengths at the physical quark mass.  
  }\label{fig:InvScatt_run}
\end{center}
\end{figure}

Our results for the chiral extrapolation of the deuteron
binding energy and the inverse S-wave scattering lengths at NLO and
N$^2$LO are visualised in Figs.~\ref{fig:BErun} and \ref{fig:InvScatt_run}.
In these figures, $m_{q0}$ denotes the physical value of the light quark mass --- note
that $m_q/m_{q0}$ is to a very good approximation equal to $M_\pi^2/M_{\pi,\,\textsl{physical}}^2$.
In these calculations, we also took into account the implicit quark mass
dependence in the two-pion exchange potential induced by $g_A$ and
$F_\pi$ which is, strictly speaking, a higher-order effect. We will
comment on the size of these higher-order contributions in the next
section. Also, as already explained above, we follow here our general
strategy and use the most accurate available information regarding
the $m_q$-dependence of $m$ and, especially, of $g_A$ coming, in
particular, from lattice QCD simulations rather than
sticking to the strict chiral expansion at a given order. Note also
that within the LS framework not all contribution to the quark mass
dependence are generated, but this effect is well within the error bands
discussed later.

We observe the opposite trends in the $^1S_0$ and $^3S_1$ channels when
changing the value of the quark mass. In particular, the interaction
between the nucleons is found to become more attractive in the $^1S_0$
channel with increasing the light quark masses while more repulsive in the $^3S_1$ partial wave.  
The obtained results do not exclude the possibility of
a bound spin-singlet state at sufficiently large quark masses.  The
deuteron is found to remain bound for all values of the quark masses
considered. 
Notice further that our results indicate that the infrared
limit cycle proposed in Ref.~\cite{Braaten:2003eu} (see also Ref.~\cite{Epelbaum:2006jc}) is
unlikely to emerge in the range of the quark masses considered in the
present analysis.  
A detailed comparison of our findings with the available
calculations will be presented in the next section.  

Let us now address the theoretical uncertainty of our calculations. 
It is comforting to see that the results for the quark mass dependence of the deuteron binding
energy and the S-wave scattering lengths calculated at NLO and N$^2$LO 
are consistent with each other. The NLO and N$^2$LO bands resulting
from the cutoff variation as described above overlap except for large
quark masses in the spin-triplet channel. 
This is in line with the observation that the cutoff variation at NLO 
underestimates the true theoretical uncertainty at this order
since the width of the bands at both NLO and N$^2$LO measures the impact of the 
neglected order-$Q^4$ contact interactions.   
A more complete discussion of the theoretical uncertainties of the
calculated $NN$ observables will be given in \Sec{sec:discuss}. 
Notice further that the chiral extrapolations become rather uncertain
for small quark masses which, at first sight, might appear counterintuitive.
This, however, has to be expected given that the LECs accompanying contact
interactions are fit to experimental data which obviously correspond 
to the physical quark masses. In addition, one should also
keep in mind that we do not rely here on the chiral expansion of the
short-range forces, contrary to
Refs.~\cite{chiralextrapol_bonn,Mondejar:2006yu}. Our approach,
utilizing resonance saturation and the $K$-factors for heavy-meson masses, 
cannot be expected to yield reliable results 
at low quark masses where short-range contributions 
non-analytic in quark masses, which are not explicitly taken into account
in our calculations, should play an important role. 

Finally, we list  in Table~\ref{tab:individual} the individual
contributions of various mechanisms to the dimensionless quantities
$K_{\rm deut}^q$ and $K_{\rm a, \; I}^q$ with $I=\{1S0, \; 3S1 \}$
defined as
\beq
K_{\rm deut}^q = \frac{\delta E_{\rm deut}/\delta m_q}{E_{\rm
    deut}/m_q}\,, \quad \quad
K_{a, \rm \; I}^q = \frac{\delta  a_{\rm I}  /\delta m_q}{
  a_{\rm I}  / m_q }\,.
\eeq
\begin{table*}[tb] 
\begin{center}
\begin{tabular}{|rcc|cc rccc rccc cc rccc rccc cc rcccr|}
\hline \hline
  &&&&& \multicolumn{8}{c}{$K_{a, \rm  \; 1S0}^q$} &&&
  \multicolumn{8}{c}{$K_{a, \rm \; 3S1}^q$}  &&&  
\multicolumn{5}{c|}{$K_{\rm deut}^q$}   \\
  &&&&&  NLO &&&& N$^2$LO &&&&&& NLO &&&& N$^2$LO &&&&&& NLO &&&& N$^2$LO \\
\hline
$V_{1\pi} + V_{\rm cont}^{(0)}$ &&&&&  
$ {0.2}  {{+0.1} \atop {-0.6}}$ &&&&   $ {-0.8}  {{+0.8} \atop {-0.5}}$
&&&&&&   
$ {0.36}  {{+0.09} \atop {-0.03}}$ &&&&  $ {0.54}  {{+0.00} \atop {-0.05}}$ 
&&&&&& 
$ {-0.87}  {{+0.06} \atop {-0.22}}$ &&&& $ {-1.28}  {{+0.12} \atop {-0.0}}$\\
${} + V_{2\pi} + V_{\rm cont}^{(2)}$  &&&&&  
$ {-1.3}  {{+2.7} \atop {-0.3}}$ &&&&   $ {1.8}  {{+1.5} \atop {-1.5}}$
&&&&&&   
$ {0.24}  {{+0.06} \atop {-0.34}}$ &&&&  $ {0.43}  {{+0.08} \atop {-0.11}}$ 
&&&&&& 
$ {-0.66}  {{+0.80} \atop {-0.13}}$ &&&& $ {-1.11}  {{+0.27} \atop {-0.19}}$\\
${} + m$ (LS eq.) &&&&&  
$ {-0.6}  {{+2.6} \atop {-0.2}}$ &&&&   $ {2.3}  {{+1.6} \atop {-1.5}}$
&&&&&&   
$ {0.13}  {{+0.05} \atop {-0.33}}$ &&&&  $ {0.32}  {{+0.08} \atop {-0.10}}$ 
&&&&&& 
$ {-0.41}  {{+0.76} \atop {-0.13}}$ &&&& $ {-0.86}  {{+0.24} \atop {-0.18}}$\\
\hline \hline
  \end{tabular}
\vspace{0.1cm}
\caption{Various contributions to  $K_{a, \rm  \; 1S0}^q$, $K_{a, \rm
    \; 3S1}^q$ and $K_{\rm deut}^q$. The numbers correspond to the
  third cutoff combination in Eq.~(\ref{cutoffs_3N}) while the errors
  result from the cutoff variations. 
\label{tab:individual}}
\end{center}
\end{table*}
We observe a reasonable convergence pattern in the triplet channel 
with the main effect coming from the LO terms in the potential 
and the contributions due to NLO$+$NNLO terms 
and the quark mass dependence of the nucleon mass being considerably
smaller. The much larger uncertainty in the singlet channel can be
explained by the known feature that the one-pion exchange yields only a minor 
contribution to the $^1S_0$ phase shift. The dominant effects emerge
from two-pion exchange and shorter-range terms whose quark mass
dependence is less constrained than the one associated with the longest-range one-pion
exchange potential.

\section{Discussion}
\label{sec:discuss}

We are now in the position to discuss in some detail the theoretical
uncertainty of our calculations. Its main sources are due to
\begin{enumerate}
\item
The uncertainty associated with the chiral extrapolation of the nucleon
mass $m$ as well as  the axial coupling $g_A$
and the pion decay constant $F_\pi$, which impact the $m_q$-dependence of the long-range    
interactions.
\item
The uncertainty due to truncating the chiral expansion of the
potential at N$^2$LO. 
\item
The uncertainty associated with the resonance saturation hypothesis for
short-range operators and the employed quark mass dependence of the
heavy-meson properties. 
\end{enumerate}

The chiral dependence of $F_\pi$ is well reproduced at
the leading-loop order in ChPT, so that the associated uncertainty has a much
smaller impact on the two-nucleon observables considered in this
work as compared to other sources. On the other hand, the chiral
expansion of $g_A$ is known to converge slowly. As explained in
\Sec{sec:ga}, we incorporate the  order-$Q^4$ counterterm and
use the lattice QCD result at $M_\pi = 353\,$MeV as an input to tune
the corresponding LEC. This allows us to obtain a realistic description of
the quark mass dependence of $g_A$.  
\begin{figure}[t!]
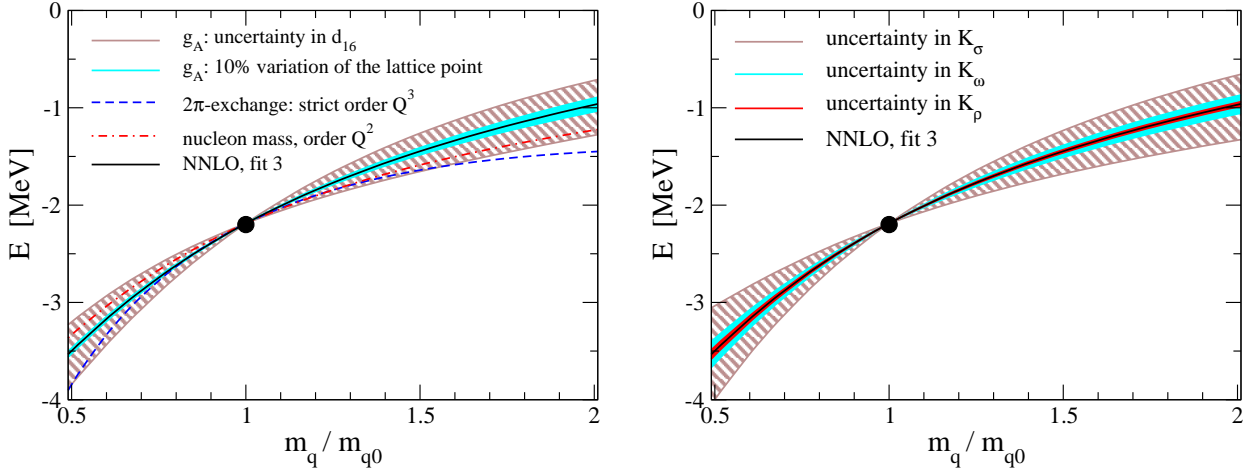

\begin{center}
\epsfig{file=errors_long_range2_mod.eps, width=0.48\textwidth}
\hfill
\epsfig{file=errors_short_range_true_mod.eps, width=0.48\textwidth}
\caption{Quark mass dependence of the deuteron binding energy  at NNLO, fit 3 (solid lines). 
Left panel: theoretical uncertainty associated with the
 quark mass dependence of the long-range interactions and the nucleon
 mass. The hatched
 band corresponds to the variation of $\bar d_{16}$ in
 the range $\bar d_{16}= -0.91$ to $-2.61$ GeV$^{-2}$ (see
~\cite{Fettes:2000fd}).  The light shaded band results from a $10\%$ variation
 of the lattice point used to fix the order-$Q^4$ counter term in
 Eq.~(\ref{ga}). Finally, 
the long-dashed-dotted line shows the effect of neglecting
 the quark mass dependence of $g_A$ and $F_\pi$ in the $2\pi$-exchange
 potential while the short-dashed-dotted line shows the effect of
 using the order-$Q^2$ rather than order-$Q^3$ expression (\ref{mnucl}) for the
 chiral extrapolation of the nucleon mass. 
Right panel: theoretical uncertainty induced by the
errors in the quark mass dependence of the heavy meson
masses according to \Tref{tab:Kestimates}. 
  }\label{fig:errors}
\end{center}
\end{figure}
We check the robustness of this procedure by allowing for a 
$10\%$ variation of the lattice point. As visualized in the left panel
of Fig.~\ref{fig:errors}, this induces a shift in the binding
energies which
is considerably smaller than our estimated theoretical
uncertainty, cf.~Fig.~\ref{fig:BErun}. On the other hand, the uncertainty in the
determination of the LEC $\bar d_{16}$,  $\bar d_{16}= -0.91$ to $-2.61$ GeV$^{-2}$
~\cite{Fettes:2000fd}, leads to a sizeable spread  which is comparable with the
 one emerging from the cutoff variation. This can be expected
 since the value of $\bar d_{16}$ influences the shape of the quark
 mass dependence of $g_A$ (larger in magnitude values lead to a more
 flat behavior), see Fig.~\ref{fig:ga}. 

The uncertainty due to truncating the chiral expansion for the
potential at N$^2$LO was already roughly estimated by the cutoff variations,
see Fig.~\ref{fig:BErun}. As an additional check, we calculated the
impact of the $M_\pi$-dependence of $g_A$ and $F_\pi$ in the
$2\pi$-exchange potential which nominally starts to contribute at
N$^3$LO (i.e. order  $Q^4$) which is beyond the accuracy of this
work. The size of this effect is given by the difference between the
solid and long-dashed-dotted lines in the left panel of
Fig.~\ref{fig:errors} and is indeed within the estimated theoretical
accuracy  at N$^2$LO. Similarly, it is, strictly speaking,
sufficient to use the order-$Q^2$ rather than order-$Q^3$ expression 
for the $M_\pi$-dependence of the nucleon mass at N$^2$LO. The induced
difference agrees well with the estimations based on
dimensional arguments and is within the accuracy of our calculation, 
see the short-dashed-dotted line in the left
panel Fig.~\ref{fig:errors}. 

Presumably, the most important source of uncertainty is due to 
the quark mass dependence of the contact interactions. While
resonance saturation itself seems, at least in principle, to provide a fairly accurate way to
relate the chiral extrapolations of the short-range terms to the
properties of heavy mesons which are easier accessible theoretically,
it is difficult to estimate the theoretical uncertainty associated
with this procedure. We, therefore, restrict ourselves to propagating
the errors in the $K^q$-factors for the heavy meson masses, see
\Tref{tab:Kestimates}, through our analysis. These errors
turn out to be strongly
magnified due to the large cancellations between the contributions of
the $\rho$ and $\omega$ mesons, see Fig.~\ref{fig:LECrun}. 
The resulting 
uncertainty in the deuteron binding energy appears to be comparable to the
one emerging from the cutoff variation and the chiral extrapolations
of $g_A$, see Fig.~\ref{fig:errors}. 
We further emphasize that using the linear
approximation in terms of the $K$-factors for the quark mass
dependence of the heavy mesons is yet another approximation
(if one goes sufficiently far away from the physical
point). It is, however, irrelevant as long as one is only interested
in the $K$-factors and can be easily avoided if necessary.  

The final results for the $K$-factors of the deuteron binding energy and the corresponding
scattering lengths read
\beq 
\boxed{
K_{a, \rm  \; 1S0}^q =  {2.3}  {{+1.9} \atop {-1.8}}\,, \quad \quad
K_{a, \rm  \; 3S1}^q =  {0.32}  {{+0.17} \atop {-0.18}}\,, \quad \quad
K_{\rm deut}^q =  {-0.86}  {{+0.45} \atop {-0.50}}\,, }
\label{Kfacs}
\eeq
where the numbers are given for the third cutoff combination at
N$^2$LO and the central values of $K_{\sigma, \rho}^q$
listed in Table~\ref{tab:Kestimates}. The 
theoretical uncertainties due to truncating higher order terms
(estimated by the cutoff variation), the uncertainty in $\bar d_{16}$,
the lattice calculation of $g_A$, and the errors in $K_{\sigma,
  \rho}^q$ are added in quadrature. 

The results given in Eq.~(\ref{Kfacs}) can be compared with the ones
of Ref.~\cite{chiralextrapol_bonn},   carried out at NLO in chiral EFT  
\beq 
K_{a, \rm  \; 1S0}^q =  5 \pm 5\,, \quad \quad
K_{a, \rm  \; 3S1}^q =  1.1 \pm 0.6\,, \quad \quad
K_{\rm deut}^q = -2.8 \pm 1.2 \,, 
\label{KfacsOurOld}
\eeq
where the numbers are inferred from  Figs. 11 and 12 of this work. 
A more conservative error estimation taking into account a larger
variation in the LEC $\bar d_{16}$ and in the quark mass dependence of
the lowest-order spin-triplet contact interaction was performed in
Ref.~\cite{bonn} leading to  
\beq 
K_{a, \rm  \; 3S1}^q =  1.1 \pm 0.9\,, \quad \quad
K_{\rm deut}^q = -2.9 \pm 1.8 \, .
\label{KfacsOurOldConcerv}
\eeq
As already pointed out before, the major differences between the
present analysis and the one of \cite{chiralextrapol_bonn,bonn} are in
using a more realistic result for the chiral expansion of $g_A$,
employing resonance saturation to describe the quark mass dependence
of contact interactions and extending the calculations to N$^2$LO.  
These improvements have allowed us to determine the values for
the $K$-factors with higher accuracy. 

The results for both $^1S_0$ and $^3S_1$ channels given above are
consistent with the chiral EFT values  calculated 
in Ref.~\cite{seattle} using the Kaplan-Savage-Wise
approach,  
\beq 
K_{a, \rm  \; 1S0}^q =  2.4 \pm 3.0 \,, \quad \quad
K_{a, \rm  \; 3S1}^q =  3.0 \pm 3.5 \,, \quad \quad
K_{\rm deut}^q =  -7   \pm 6 \,,
\label{KfacsKSW}
\eeq
where the numbers correspond to Figs. 1, 2 and 4  ($\eta
= 1/5$) of that work. 

More recently, attempts have been made to combine chiral EFT with 
lattice-QCD calculations. In particular, the NPLQCD collaboration
has determined the regions for the S-wave scattering lengths
consistent with their lattice data $a_{1S0}= 0.63 \pm 0.50 (5$-$10)$ fm
and $a_{3S1}=0.63 \pm 0.74 (5$-$9)$ fm obtained at $M_\pi = 353.7 \pm
2.1$ MeV \cite{Beane:2006mx}.  Assuming the validity of the employed chiral EFT frameworks 
in the considered range of pion masses, they determined the following
constraints for the allowed regions of the $K$-factors (the numbers below are extracted from
Figs. 3, 4 of Ref.~\cite{Beane:2006mx}):   
\beq
\label{NPLQCD1}
K_{a, \rm  \; 1S0}^q  \lesssim -4 \;\;  \cup \; \; K_{a, \rm  \;
  1S0}^q  \gtrsim 2
\eeq
based on the Weinberg approach and 
\beq
\label{NPLQCD2}
K_{a, \rm  \;
  1S0}^q  \gtrsim 6 \,,  \quad \quad    -5 \lesssim K_{a, \rm  \; 3S1}^q
\lesssim -0.2
\;\;  \cup \; \; 
 0.3 \lesssim K_{a, \rm  \; 3S1}^q
\lesssim 9\,,
\eeq
using the chiral EFT formulation of Ref.~\cite{Beane:2001bc}.
For a more recent extrapolation of the NPLQCD numbers to
physical quark masses see Ref.~\cite{chen}. 
Very recently, a similar analysis has been carried out using the
framework of chiral EFT with dibaryon fields~\cite{soto_new}
yielding $K_{a, \rm \;
  3S1}^q \sim - 0.4$, see Fig.~9 of that work, and 
a positive value for $K_{a, \rm \; 1S0}^q$ (which we were unable to
infer from that figure).  While this result for $K_{a, \rm \;
  3S1}^q$ disagrees with our findings, it is difficult to draw
conclusions since Ref.~\cite{soto_new} does not provide an
estimate of the
theoretical uncertainty associated with using the lattice-QCD results
at large values of the pion mass as input in the calculations. The
same comment 
also applies to the results of Ref.~\cite{Beane:2006mx} given in
Eqs.~(\ref{NPLQCD1}),  (\ref{NPLQCD2}) as well as the results of Ref.~\cite{chen}.

Last but not least, it is comforting to see that our results are in a good agreement with
the  ones of Ref.~\cite{victorandbob}, where the value $K_{\rm deut}^q
=-0.75$, $-0.84$ and  
$-1.39$  is reported for three different models of the two-nucleon
potentials. Even more important is, however, that we are able to
carefully estimate the theoretical uncertainty for this quantity.

\section{Consequences for heavier nuclei}
\label{sec:heavy_dep}

So far we focused on the two--nucleon system, however,  the quark
mass dependences of  $^3$He and $^4$He are also relevant for BBN.
In order to estimate the impact of the quark mass dependences of the deuteron
binding energy and the $^1S_0$ scattering length on BBN, we here use the 
methods of   Ref.~\cite{platterbedaque} --- actually, what we have provided in
the previous sections is an update of the input used in
Ref.~\cite{platterbedaque}, which basically came from Ref.~\cite{chiralextrapol_bonn}.

The quark mass dependences of the binding energies of  $^3$He and $^4$He can 
be calculated from 
\begin{equation}
K_{^A{\rm He}}^q = K_{a, \rm  \; 1S0}^q K_{^A{\rm He}}^{a, \rm  \; 1S0}
+K_{\rm deut}^qK_{^A{\rm He}}^{\rm deut} \ ,
\label{HeK}
\end{equation}
where the $K_{^A{\rm He}}^x=x/E_{^A{\rm He}}(\delta E_{^A{\rm
    He}}/\delta x)$ were defined in analogy to the quantities given above.
In Ref.~\cite{platterbedaque} the coefficients were calculated from pionless
EFT. They read, including the uncertainties quoted in Ref.~\cite{platterbedaque}
\begin{eqnarray} \nonumber
K_{^3{\rm He}}^{a, \rm  \; 1S0} = 0.12\pm 0.01 \quad , \qquad K_{^3{\rm He}}^{\rm
  deut}=1.41\pm 0.01 \ ; \\
\label{eq:bedaque}
K_{^4{\rm He}}^{a, \rm  \; 1S0} = 0.037\pm 0.011 \quad , \qquad K_{^4{\rm He}}^{\rm
  deut}=0.74\pm 0.22 \ ; 
\end{eqnarray}
From this we get using the uncertainties for the $K$--factors as given
in Eq.~(\ref{Kfacs})
\begin{equation}
\label{eq:K_He}
K_{^3{\rm He}}^q = -0.94\pm 0.75 \quad , \qquad K_{^4{\rm He}}^q = -0.55\pm 0.42 \ ,
\end{equation}
where the uncertainties of Ref.~\cite{platterbedaque} and those quoted in
Eq.~(\ref{Kfacs}) were added in quadrature.
Note that there has been a recent lattice study of the deuteron, $^3$He and 
$^4$He at a pion mass of 510~MeV and various lattice sizes \cite{Yamazaki:2012hi}. 
As this pion mass is rather large, we refrain from trying to extract the
corresponding $K$-factors from that work. In the future,
 when such studies
become available at lower values of $M_\pi$, they will provide valuable
constraints on the quark mass dependence of nuclear binding energies.
Note further that the $K$-factor for $^4$He is consistent for the central
value obtained from a recent nuclear lattice calculation using the same 
scattering lengths $K$-factors, $K_{^4{\rm He}}^q = -0.32$ \cite{Epelbaum:2012iu}.

\section{Impact on BBN}
\label{sec:bbn}

In \Tref{tab:bbn_response} we present our calculated BBN response matrix. The
quantities presented are the linear dependences of calculated primordial
abundances $Y_a$ to small variations of nuclear binding energies and
scattering lengths $X$, $\partial\ln Y_a/\partial\ln X$. They were obtained
using the methods and codes (modified from the publicly available Kawano
code~\cite{kawano92fermilab}) described in~\cite{berengut10plb}. Updated
reaction rates are taken from
Refs.~\cite{angulo99npa,cyburt04prd,ando06prc,cyburt08prc},
see~\cite{berengut10plb} for details. Unlike in previous studies, we have
separated the effect of $B_\textrm{deut}$ from the singlet scattering length
$a_s$. The scattering length affects the rate of the reaction $n(p,d)\gamma$
via the equation (see, e.g.~\cite{segre77book})
\[
\left< \sigma v\right> \sim (B_\textrm{deut})^{5/2}/\epsilon_v
\]
where $\epsilon_v$ is the position of the singlet virtual level. The baryon-to-photon ratio $\eta=6.19\,(15)\E{-10}$ is taken from the latest WMAP7 data~\cite{larson11apjss}.

\begin{table}[tb]
\begin{ruledtabular}
\begin{tabular}{lrrrrr}
X     &     d & \HeT & \HeF & \LiSix & \LiS \\
\hline
$a_s$ & $-$0.39 &  0.17 &  0.01 & $-$0.38 & 2.64 \\
$B_\textrm{deut}$
      & $-$2.91 & $-$2.08 &  0.67 & $-$6.57 & 9.44 \\
$B_\textrm{trit}$
      & $-$0.27 & $-$2.36 &  0.01 & $-$0.26 & $-$3.84 \\
$B_{\HeT}$
      & $-$2.38 &  3.85 &  0.01 & $-$5.72 & $-$8.27 \\
$B_{\HeF}$
      & $-$0.03 & $-$0.84 &  0.00 & $-$69.8 & $-$57.4 \\
$B_{\LiSix}$
      &  0.00 & 0.00  &  0.00 &  78.9 & 0.00 \\
$B_{\LiS}$
      &  0.03 & 0.01  &  0.00 &  0.02 & $-$25.1 \\
$B_{\BeS}$
      &  0.00 &  0.00 &  0.00 &  0.00 & 99.1 \\
$\tau$
      &  0.41 & 0.14  & 0.72 & 1.36 & 0.43 \\
\end{tabular}
\end{ruledtabular}
\caption{\label{tab:bbn_response} BBN response matrix $\partial\ln
  Y_a/\partial\ln X$ at 
 baryon-to-photon ratio $\eta=6.19\E{-10}$. The $Y_a$ are the number ratios of primordial isotope abundances to hydrogen, except for \HeF\ which is the mass ratio $\HeF/\textrm{H}$. }
\end{table}

Final sensitivities of primordial abundances $Y_a$ to variation of $m_q$, are
obtained by combining the results in \Tref{tab:bbn_response} with
Eqs.~(\ref{Kfacs}, \ref{eq:K_He}) using \beq \frac{\delta \ln Y_a}{\delta\ln
  m_q} = \sum_{X_i} \frac{\partial\ln Y_a}{\partial \ln X_i} \, K^q_{X_i} \,.
\eeq Error estimates must be performed carefully, since the $K^q_{X_i}$ are
derived from the same sources via Eq.~\eref{HeK}. We have taken the correlations
in errors into account when deriving our final sensitivities. The
uncertainties of Eq.~\eref{eq:bedaque} are also correlated to some extent, but
since they are small anyway we may neglect this correlation. The final
sensitivities of primordial abundances to quark mass variation are given in
\Tref{tab:bbn_results}, along with final values of quark mass variation at the
time of big bang nucleosynthesis extracted from comparison of observed and
calculated primordial abundances.

\begin{table*}[tb]
\begin{ruledtabular}
\begin{tabular}{ccccc}
$Y_a$ & obs. & calc. & ${\delta \ln Y_a}/{\delta\ln m_q}$ & $\delta m_q/m_q$ \\
\hline
[deut/H]  & $2.82\,(21)\E{-5}$ & $2.49\,(17)\E{-5}$ & $3.9\,(3.4)$ & $0.034\,(42)$ \\
$\HeF\ (Y_p)$ & $0.249\,(9)$ & $0.2486\,(2)$ & $-0.56\,(34)$ & $-0.003\,(65)$ \\
\end{tabular}
\end{ruledtabular}
\caption{\label{tab:bbn_results} Extracted values of quark mass variation, $\delta m_q/m_q$, during BBN from comparison of observed and calculated primordial abundances, $Y_a$. Observed values are taken from the Particle Data Group review~\cite{beringer12prd} and calculated values from~\Cite{cyburt08jcap}.}
\end{table*}

We see from \Tref{tab:bbn_results} that the deuterium and \HeF\ data give
consistent limits on $\delta m_q/m_q$. Taking a weighted average of the two
results gives $\delta m_q/m_q = 0.02\pm0.04$, our final result.

\section{Effect of the Neutron Lifetime}
\label{sec:tau}

Our limit $\delta m_q/m_q = 0.02\pm0.04$ at first appears much weaker than the
limit derived by Bedaque \etal~\cite{platterbedaque}, who obtained $-1\%
\lesssim \delta m_q/m_q \lesssim 0.7\%$, although our input for the
two--nucleon parameters is more accurate. The origin of
the difference is that, contrary to this work as well as previous
works~\cite{dent07prd,victorandbob,berengut10plb},
 in Ref.~\cite{platterbedaque} a quark mass dependence of
the neutron lifetime, $\tau$, was included.
Since essentially all free neutrons at the onset of BBN
end up in \HeF\ nuclei, and changing $\tau$ changes the neutron-to-proton
number ratio at BBN, the \HeF\ abundance is quite sensitive to the
neutron decay rate. Therefore it is worthwhile to examine the assumptions made in
\Cite{platterbedaque} in more detail.

Neutron beta decay becomes possible as a consequence of a non--vanishing proton--neutron
mass difference, which is non--zero due to an apparent violation of the
isospin symmetry in the Standard Model caused by  $m_u \neq m_d$ and electromagnetic
effects driven by different quark charges. One finds
\begin{equation}
\Delta m_N=m_n-m_p=\Delta m_N^{\rm str}+\Delta m_N^{\rm em} = 2 \ \mbox{MeV}
  - 0.7 \ \mbox{MeV} ,
\end{equation}
with an uncertainty of 0.3 MeV for the individual
contributions~\cite{Gasser}\footnote{The more recent extraction of
  Ref.~\cite{jerry} finds $\Delta m_N^{\rm em}=-1.3\pm 0.5$ MeV --- consistent
  within uncertainties.}.
Thus, in order to quantify how $\Delta m_N$ and thus $\tau$ changes with the quark masses, an
assumption has to be made on how the other Standard Model parameters change
--- as we will see of particular importance is the change of the electron mass
relative to the quark masses.



The neutron width $\Gamma \sim 1/\tau$  can be written as
\begin{equation}
\label{eq:Gamma}
\Gamma = \frac{(G_F\cos\theta_C)^2}{2\pi^3}\,m_e^5 \,(1+3 g_A^2)\,
         f\left(\frac{\Delta m_N}{m_e}\right)
\end{equation}
where $G_F$ is the Fermi constant, $\theta_C$ is the Cabibbo angle, $m_e$ is the electron
mass, and $g_A$ is the nucleon axial decay constant. The function
$f(\Delta m_N/m_e)$ describes the phase space and Coulomb attraction; an explicit
form is presented in \cite{platterbedaque}.

In order to proceed, \Cite{platterbedaque} made the assumption
that when $m_q$ changes, $m_u/m_d$ as well as all other Standard Model parameters
stay constant. This clearly introduces some model-dependence.
Based on this assumption one gets a very
 strong sensitivity of $\Gamma$ to $m_q$ 
via the factor $f$:
\begin{equation}
\label{eq:f}
\frac{\delta\ln\Gamma}{\delta\ln m_q}\simeq \frac{\delta\ln f(\Delta m_N/m_e)}{\delta\ln m_q}
= \left.\frac{f'}{f}\right|_{\Delta m_N/m_e} \left(\frac{\delta(\Delta m_N/m_e)}{\delta\ln
  m_q}\right)=10.4\pm 1.5 \, ,
\end{equation}
where ${f'}/{f}$ is  numerically determined to be 2.57 at the physical value of
$\Delta m_N/m_e$. Thus, the large sensitivity
to the variation of the quark mass,  found in \Cite{platterbedaque}, comes
directly from model-dependent
assumptions, which allow one to write 
$$\delta(\Delta m_N/m_e)/\delta\ln
  m_q=\Delta m_N^{\rm str}/m_e=4 \, .$$
One the other hand, had  one assumed that $m_u-m_d$ is independent of the
quark mass, the value found for $\delta\ln\Gamma/\delta\ln m_q$ would  be
significantly smaller. 

In general, how possible changes of fundamental parameters are interrelated depends
on the model assumed for the physics beyond the Standard Model. In fact,
the relation $m_u/m_d =$ constant emerges naturally from a scenario where all elementary
particle masses are proportional to a varying Higgs vacuum expectation value,
$v$ (relative to $\Lambda_\textrm{QCD}$), but the gauge and Yukawa couplings
are independent of it. However, in this case $m_e$ and the mass of the weak
gauge boson, $m_W$, are also proportional to $v$. One finds, therefore, that
$\Delta m_N^{\rm str}/m_e$ has no dependence on $v$, and the sensitivity of
$f$ comes from the variation in $m_e$ relative to the 
electromagnetic component of $\Delta m_N$:
\begin{equation}
\frac{\delta\ln f(\Delta m_N/m_e)}{\delta\ln v}
= \left.\frac{f'}{f}\right|_{\Delta m_N/m_e} \left(\frac{-\Delta m_N^{\rm em}}{m_e}\right)
= 3.6\pm 1.5 \,.
\end{equation}
Under our assumption that $\delta\ln m_W =
\delta\ln m_e = \delta\ln m_q = \delta\ln v$ (i.e. all masses vary with the
Higgs VEV), and noting that $G_F\sim 1/m_W^2$, from \eref{eq:Gamma} we obtain
\begin{equation}
\label{eq:dgamma_dv}
\frac{\delta\ln\Gamma}{\delta\ln v}
= 1 + \frac{3g_A^2}{1+3g_A^2}\frac{\delta\ln g_A^2}{\delta\ln v} + \frac{\delta\ln f(\Delta m_N/m_e)}{\delta\ln v}
= 4.8\pm 1.5 \ ,
\end{equation}
where we used the quark mass dependence of $g_A^2$ discussed in Sec.~\ref{sec:ga}, which
gives 
\begin{equation}
\frac{\delta\ln g_A^2}{\delta\ln v} = 0.2\pm 0.1 \ .
\end{equation}
The final sensitivity to neutron decay rate, ${\delta\ln\tau}/{\delta\ln v} =
-4.8$, is reduced by around a factor of two compared
with~\cite{platterbedaque}, but it is still very large even within our model.
Note, ${\delta\ln\tau}/{\delta\ln v}=-4.9$, may also be obtained
using Table~IV of Dent~\etal~\cite{dent07prd} under the same assumptions, and
a value of $-4.8$ can be extracted from Eq.~(7) of Ref.~\cite{coc07prd}.
Multiplying ${\delta\ln\tau}/{\delta\ln v}$ by the sensitivity coefficients
$\delta\ln Y_a/\delta\ln\tau$ presented in \Tref{tab:bbn_response} and adding
the binding energy and scattering length sensitivities presented in
\Tref{tab:bbn_results} (using our assumption $\delta\ln m_q = \delta\ln v$),
we obtain
\begin{equation}
\frac{\delta\ln Y_{deut}}{\delta\ln v} = 1.9\,(3.4)\ , \qquad
\frac{\delta\ln Y_{\HeF}}{\delta\ln v} = -4.0\,(0.3) \ .\\
\end{equation}
In this model, the $\tau$ sensitivity reduces the deuterium sensitivity to
$v$ by a factor of two, and entirely dominates the \HeF\ sensitivity. The
final limits on Higgs VEV variation from deuterium and helium abundances are
$\delta v/v = 0.07\pm0.13$ and $\delta v/v = 0.000 \pm 0.009$, respectively.
However, we should put here the disclaimer that, although we included the
effects of isospin violation in the evaluation of the neutron beta decay, 
all few-nucleon calculations were done imposing isospin symmetry.
One expects these effects to be small compared to that from
 neutron beta decay; still, for consistency such effects will have to 
be included in the future.

Therefore, imposing the model that all masses
scale with $v$ and including the variation of $\tau$, then as expected the
results are  entirely dominated by the \HeF\ data and the limits are rather
tight, $|\delta v/v| < 0.9\%$. Finally, we note that this result does not
include all possible mass variations: for example, in
\Cite{dent07prd} the dependences of primordial abundances on $m_e$ are calculated
assuming $\tau$ constant (that is, the $m_e$ dependence that does not come via
$\tau$ variation). In comparison with the dependence from $\tau$ these are
rather small, $\delta\ln Y_a/\delta\ln m_e = -0.16$ and $-0.71$ for deuterium
and \HeF, respectively. Including this effect would marginally tighten our
limits on $\delta v/v$.

\section{Summary}

We have presented a systematic study of the impact of quark-mass variations
on properties of the two--nucleon system based on chiral perturbation theory combined
with non-perturbative techniques. Since the approach is based on an effective
field theory a reliable error estimate becomes feasible --- a clear advantage
compared to purely phenomenological studies as reported, e.g., in 
Refs.~\cite{victoredward,craigs,craigs2,adelaide,flambaum03arxiv,flambaum04prd}.
We include the uncertainties from the quark mass dependence of
meson
masses as well as those from the treatment of the $NN$ interaction derived from chiral
perturabtion theory. Here the heavy
(heavier than the pion) meson masses enter in the expressions for
four-nucleon contact
terms with strengths fixed from a resonance saturation hypothesis --- the 
uncertainty induced by this method is not included. However, support for this procedure
comes from
the apparent quantitative success of the resonance saturation for four-nucleon
contact interactions.
 Especially we find
\beq 
K_{a, \rm  \; 1S0}^q =  {2.3}  {{+1.9} \atop {-1.8}}\,, \quad \quad
K_{a, \rm  \; 3S1}^q =  {0.32}  {{+0.17} \atop {-0.18}}\,, \quad \quad
K_{\rm deut}^q =  {-0.86}  {{+0.45} \atop {-0.50}}\,, 
\eeq
where the uncertainties are significantly reduced compared to the numbers derived from earlier studies 
within effective field theory~\cite{chiralextrapol_bonn,chiralextrapol_seattle}.
The numbers for the $K$-factors presented in this work are the necessary input for studies that address 
the quark mass dependence of nuclear properties. These studies allow one to
quantify how much fine-tuning is necessary amongst the Standard Model
parameters to allow life to develop --- for the most recent developments in
this respect see Ref.~\cite{Epelbaum:2012iu}.

 From the given $K$--factors we  derived the quark-mass
dependence of helium nuclei using the techniques of Ref.~\cite{platterbedaque}.
Additionally we have presented a new response
matrix of calculated primordial abundances to variations in nuclear binding
energies and the scattering length of the important two-nucleon $^1S_0$
channel. Combining these we have derived a stringent limit on the quark-mass
variation  at the time of big bang
nucleosynthesis of $\delta m_q/m_q = 0.02\pm0.04$.

In previous phenomenological studies
the  bounds derived from the deuteron and $^4$He abundances are, e.g., $0.009\pm 0.19$ 
and $-0.005\pm 0.038$, respectively, from Ref.~\cite{victorandbob}
and  $-0.002\pm 0.037$ 
and $0.012\pm 0.011$, respectively, from  Ref.~\cite{berengut10plb}.
The uncertainties are significantly smaller  compared to
ours, since in these works no attempt
was made to quanitify the theoretical uncertainty --- in
Ref.~\cite{victorandbob} it is stated that the uncertainty is expected to be
of the order of a factor of 2. 
In this sense, although the uncertainty of our work seems larger, still the bound
 derived is more robust, since a careful uncertainty estimate was done.

In Ref.~\cite{platterbedaque} a range $-1 \%\leq \delta m_q/m_q \leq 0.7\%$ is
quoted for the quark mass variation allowed by BBN. In their calculation, the sensitivity
is dominated by variation of the neutron lifetime, $\tau$, which strongly affects the
\HeF\ abundances which are well-constrained observationally.
We have shown that this calculation is based on a  model-dependent
assumption. 
 Under the
reasonable assumption that all elementary particle masses are proportional to $v$, the
Higgs vacuum expectation value,
the sensitivity of $\tau$ to $v$ is shown to be quite large (although only half that
found by \cite{platterbedaque}) and we obtain a limit $|\delta v/v| < 0.9\%$,
which within this model translates into $|\delta m_q/m_q| < 0.9\%$.

In Refs.~\cite{victorandbob,berengut10plb} also bounds are derived
from the \LiS\ abundance that are different from zero, namely $\delta
m_q/m_q = 0.016\pm0.005$ 
and  $\delta m_q/m_q = 0.013\pm0.002$, respectively. 
Also here the uncertainties do not include the theoretical uncertainty of the
input quantities. 
In our work we have not included the \LiS\ abundance for several
reasons. Firstly, the \LiS\ abundance is very sensitive to variations in
\BeS\ and \LiS\ binding energies, as well as $A=5$
resonances~\cite{berengut10plb} which have not yet been calculated using the
methodology of this paper. In the future our study can be extended into this
regime, as soon as systematic studies of the quark mass dependence of heavier
nuclei are available, e.g., employing methods of nuclear lattice
calculations --- see Ref.~\cite{Epelbaum:2011md,Epelbaum:2012qn} 
and references therein. In fact, the first results within that framework for
$^4$He, $^8$Be and $^{12}$C can be found in  \cite{Epelbaum:2012iu}.
Secondly, the discrepancy between theory and
observation is a factor of $\sim3$ which can make nonlinear effects
important. Lastly, the observational status and interpretation of the `Spite plateau' of lithium
abundances in Pop. III stars is still uncertain (see, e.g.,
Refs.~\cite{beringer12prd,new1,new2} 
and references therein).

\begin{acknowledgments}
C. H. acknowledges the hospitality and the funding 
by UNSW where a part of this work was carried out. J. N. acknowledges funding by Fundaci\'on Ram\'on Areces.
This work is supported in part
 by the Spanish Research contracts
FPA2007-29115-E/, FPA2011-27853-C02-02, and FPA2008-05287-E/FPA "Quark
Masses and Hadron Physics (From Quarks to life)'',
 by the Australian Research Council, by the DFG and the NSFC through
funds provided to the Sino-German CRC 110 ``Symmetries and
the Emergence of Structure in QCD'', by the HGF through funds
provided to the Nuclear Astrophysics Virtual Institute NAVI (VH-VI-417),
by the BMBF (06BN9006), by the EU via ERC project 259218 NuclearEFT  and  the EU FP7 HadronPhysics3 project.

\end{acknowledgments}

\end{document}